\documentclass{article}
\usepackage{amsthm, amssymb, amsmath}
\newtheorem{theorem}{Theorem}[section]

\newtheorem{lemma}[theorem]{Lemma}

\newtheorem{remark}{Remark}
\newtheorem{proposition}[theorem]{Proposition}
\newtheorem{corollary}[theorem]{Corollary}
\newenvironment{them}{\textbf{Theorem}\it}{}
\newtheorem{definition}[theorem]{Definition}
\huge
\begin{document}
\title{Relaxation of Solitons in Nonlinear Schr\"odinger Equations with
Potential\thanks{This paper is part of the first author's Ph.D
thesis.}}\date{}
\author{Zhou Gang\thanks{Supported by NSERC under Grant NA7901 and NSF under Grant DMS-0400526.} \ I.M. Sigal$^{\dag}$}
\maketitle 
\setlength{\rightmargin}{.1in} \centerline{\small{Department of
Mathematics, University of Toronto,
Toronto, Canada}} 
\normalsize \vskip.1in \setcounter{page}{1}
\setlength{\leftmargin}{.1in}
\setlength{\rightmargin}{.1in}
\section*{Abstract} In this paper we
study dynamics of solitons in the generalized nonlinear
Schr\"odinger equation (NLS) with an external potential in all
dimensions except for 2. For a certain class of nonlinearities such
an equation has solutions which are periodic in time and
exponentially decaying in space, centered near different critical
points of the potential. We call those solutions which are centered
near the minima of the potential and which minimize energy
restricted to $\mathcal{L}^2-$unit sphere, \textit{trapped solitons}
or just \textit{solitons}.

In this paper we prove, under certain conditions on the potentials
and initial conditions, that trapped solitons are asymptotically
stable. Moreover,
if an initial condition is close to a trapped soliton then the
solution looks like a moving soliton relaxing to its equilibrium
position. The dynamical law of motion of the soliton (i.e.
effective equations of motion for the soliton's center and
momentum)
is  close to Newton's equation but with a dissipative term due to
radiation of the energy to infinity.
\section{Introduction}
In this paper we study dynamics of solitons in the generalized
nonlinear Schr\"odinger equation (NLS) in dimension $d\neq 2$ with
an external potential $V_{h}:\ \mathbb{R}^{n}\rightarrow
\mathbb{R}$,
\begin{equation}\label{NLS}
i\frac{\partial\psi}{\partial t}=-\Delta
\psi+V_{h}\psi-f(|\psi|^{2})\psi.
\end{equation}
Here $h>0$ is a small parameter giving the length scale of the
external potential in relation to the length scale of the $V_{h}=0$
solitons (see below), $\Delta$ is the Laplace operator and $f(s)$ is
a nonlinearity to be specified later. We normalize $f(0)=0$. Such
equations arise in the theory of Bose-Einstein condensation
\footnote[1]{In this case Equation (~\ref{NLS}) is called the
Gross-Pitaevskii equation.}, nonlinear optics, theory of water waves
\footnote[2]{In these two areas $V_{h}$ arises if one takes into
account impurities and/or variations in geometry of the medium and
is, in general, time-dependent.}and in other areas.

To fix ideas we assume the potentials to be of the form
$V_{h}(x):=V(hx)$ with $V$ smooth and decaying at $\infty.$ Thus
for $h=0,$ Equation (~\ref{NLS}) becomes the standard generalized
nonlinear Schr\"odinger equation (gNLS)
\begin{equation}\label{gNLS}
i\frac{\partial \psi}{\partial t}=-\Delta
\psi+\mu\psi-f(|\psi|^{2})\psi,
\end{equation} where $\mu=V(0).$
For a certain class of nonlinearities, $f(|\psi|^{2})$ (see Section
~\ref{ExSt}), there is an interval $\mathcal{I}_{0}\subset
\mathbb{R}^{n}$ such that for any $\lambda\in \mathcal{I}_{0}$
Equation (~\ref{gNLS}) has solutions of the form
$e^{i(\lambda-\mu)t}\phi_{0}^{\lambda}(x)$ where
$\phi_{0}^{\lambda}\in \mathcal{H}_{2}(\mathbb{R})$ and
$\phi_{0}^{\lambda}>0.$ Such solutions (in general without the
restriction $\phi_{0}^{\lambda}>0$) are called the \textit{solitary
waves} or \textit{solitons} or, to emphasize the property
$\phi_{0}^{\lambda}>0,$ the \textit{ground states}. For brevity we
will use the term \textit{soliton} applying it also to the function
$\phi_{0}^{\lambda}$ without the phase factor $e^{i(\lambda-\mu)t}.$

Equation (~\ref{gNLS}) is translationally and gauge invariant.
Hence if $e^{i(\lambda-\mu)t}\phi_{0}^{\lambda}(x)$ is a solution
for Equation (~\ref{gNLS}), then so is $$
e^{i(\lambda-\mu)t}e^{i\alpha}\phi_{0}^{\lambda}(x+a),\ \text{for
any}\ a\in \mathbb{R}^{n},\text{and}\ \alpha\in [0,2\pi).$$ This
situation changes dramatically when the potential $V_{h}$ is
turned on. In general, as was shown in ~\cite{Floer,Oh1,ABC} out
of the $n+2$-parameter family
$e^{i(\lambda-\mu)t}e^{i\alpha}\phi_{0}^{\lambda}(x+a)$ only a
discrete set of two parameter families of solutions to Equation
(~\ref{NLS}) bifurcate: $e^{i\lambda
t}e^{i\alpha}\phi^{\lambda}(x),$ $\alpha\in [0,2\pi)$ and
$\lambda\in \mathcal{I}$ for some $\mathcal{I}\subseteq
\mathcal{I}_{0}$, with $\phi^{\lambda}\equiv \phi_{h}^{\lambda}\in
\mathcal{H}_{2}(\mathbb{R})$ and $\phi^{\lambda}>0$. Each such
family centers near a different critical point of the potential
$V_{h}(x).$ It was shown in ~\cite{Oh2} that the solutions
corresponding to minima of $V_{h}(x)$ are orbitally (Lyapunov)
stable and to maxima, orbitally unstable. We call the solitary
wave solutions described above which correspond to the minima of
$V_{h}(x)$ \textit{trapped solitons} or just \textit{solitons} of
Equation (~\ref{NLS}) omitting the last qualifier if it is clear
which equation we are dealing with.

The main result of this paper is a proof that the trapped solitons
of Equation (~\ref{NLS}) are asymptotically stable. The latter
property means that if an initial condition of (~\ref{NLS}) is
sufficiently close to a trapped soliton then the solution converges
(relaxes),
$$\psi(x,t)-e^{i\gamma(t)}\phi^{\lambda_{\infty}}\rightarrow 0,$$ in some
weighted $\mathcal{L}^{2}$ space to, in general, another trapped
soliton of the same two-parameter family. We also find effective
equations for the soliton center and other parameters. In this paper
we prove this result under the additional assumption that if $d>2$
then the potential is spherically symmetric and that the initial
condition symmetric with respect to permutations of the coordinates.
In this case the soliton relaxes to the ground state along the
radial direction. This limits the number of technical difficulties
we have to deal with. We expect that our techniques extend to the
general case when the soliton spirals toward its equilibrium.

In fact, we prove a result more general than asymptotic stability
of trapped solitons. Namely, we show that if an initial condition
is close (in the weighted norm
$\|u\|_{\nu,1}:=\|(1+|x|^{2})^{-\frac{\nu}{2}}u\|_{\mathcal{H}^{1}}$
for sufficiently large $\nu$) to the soliton
$e^{i\gamma_{0}}\phi^{\lambda_{0}},$ with $\gamma_{0}\in
\mathbb{R}$ and $\lambda_{0}\in \mathcal{I}$ ($\mathcal{I}$ as
above), then the solution, $\psi(t),$ of Equation (~\ref{NLS}) can
be written as
\begin{equation}\label{decom1}
\psi(x,t)=e^{i\gamma(t)}\big(e^{ip(t)\cdot
x}\phi^{\lambda(t)}(x-a(t))+R(x,t)\big),
\end{equation}
where
$\|R(t)\|_{\nu,1}\rightarrow 0$, $\lambda(t)\rightarrow
\lambda_{\infty}$ for some $\lambda_{\infty}$ as $t\rightarrow
\infty$ and the soliton center $a(t)$ and momentum $p(t)$ evolve
according to an effective equations of motion close to Newton's
equation in the potential $h^2V(a)$.

We observe that (~\ref{NLS}) is a Hamiltonian system with conserved
energy (see Section ~\ref{HaGWP}) and, though orbital (Lyapunov)
stability is expected, the asymptotic stability is a subtle matter.
To have asymptotic stability the system should be able to dispose of
excess of its energy, in our case, by radiating it to infinity. The
infinite dimensionality of a Hamiltonian system in question plays a
crucial role here. This phenomenon as well as a general class of
classical and quantum relaxation problems was pointed out by J.
Fr\"ohlich and T. Spencer ~\cite{FS}.

We also mention that because of slow time-decay of the linearized
propagator, the low dimensions $d=1,2$ are harder to handle than the
higher dimensions, $d>2.$


We refer to ~\cite{GS1} for a detailed review of the related
literature. Here we only mention results of
~\cite{Cu,Buslaev,BP2,BuSu,SW1,SW2,SW3,SW4,TY1,TY2,TY3} which deal
with a similar problem.  Like our work,
~\cite{SW1,SW2,SW3,SW4,TY1,TY2,TY3} study the ground state of the
NLS with a potential. However, these papers deal with the
near-linear regime in which the nonlinear ground state is a
bifurcation of the ground state for the corresponding
Schr\"{o}dinger operator $-\Delta+V(x).$ The present paper covers
highly nonlinear regime in which the ground state is produced by
the nonlinearity (our analysis simplifies considerably in the
near-linear case). Now, papers ~\cite{Cu,Buslaev,BP2,BuSu}
consider the NLS without a potential so the corresponding
solitons, which were described above, are affected only by a
perturbation of the initial conditions which disperses with time
leaving them free. While in our case they, in addition, are under
the influence of the potential and they relax to an equilibrium
state near a local minimum of the potential.

We formulate some open problem:

(1) Extend the results of this present paper to more general initial
conditions and to more general, probably time-dependent, potentials.

(2) Think the results of this paper with the results of ~\cite{FGJS}
on the long time dynamics of solitons.

A natural place to start here is spherically symmetric potentials
but general initial conditions. Note that for certain
time-dependent potentials the solitons will never settle in the
ground state.

As customary we often denote derivatives by subindices as in
$\phi^{\lambda}_{\lambda}=\frac{\partial}{\partial\lambda}\phi^{\lambda}$
for $\phi^{\lambda}=\phi^{\lambda}(x).$ However, the subindex $h$
signifies always the dependence on the parameter $h$ and not the
derivatives in $h.$ The Sobolev and $L^{2}$ spaces are denoted by
$\mathcal{H}^{k}$ and $\mathcal{L}^{2}$ respectively.
\section*{Acknowledgment}
We are grateful to  J. Colliander, S. Cuccagna, S. Dejak, J.
Fr\"ohlich, Z. Hu, W. Schlag, A. Soffer, G. Zhang, V. Vougalter and,
especially, V.S. Buslaev for fruitful discussions. This paper is
part of the first author's Ph.D thesis requirement.

\section{Hamiltonian Structure and GWP}\label{HaGWP}
Equation (~\ref{NLS}) is a Hamiltonian system on Sobolev space
$\mathcal{H}^{1}(\mathbb{R},\mathbb{C})$ viewed as a real space
$\mathcal{H}^{1}(\mathbb{R},\mathbb{R})\oplus
\mathcal{H}^{1}(\mathbb{R},\mathbb{R})$ with the inner product
$(\psi,\phi)=Re\int_{\mathbb{R}}\bar{\psi}\phi$ and with the
simpletic form
$\omega(\psi,\phi)=Im\int_{\mathbb{R}}\bar{\psi}\phi.$ The
Hamiltonian functional is: $$H(\psi):=\int
[\frac{1}{2}(|\psi_{x}|^{2}+V_{h}|\psi|^{2})-F(|\psi|^{2})],$$ where
$F(u):=\frac{1}{2}\int_{0}^{u}f(\xi)d\xi.$

Equation (~\ref{NLS}) has the time-translational and gauge
symmetries which imply the following conservation laws: for any
$t\geq 0,$ we have
\begin{enumerate}
 \item[(CE)] conservation of energy: $H(\psi(t))=H(\psi(0));$
 \item[(CP)]
 conservation of the number of particles: $N(\psi(t))=N(\psi(0)),$ where $N(\psi):=\int
 |\psi|^{2}.$
\end{enumerate}
To address the global well-posedness of (~\ref{NLS}) we need the
following condition on the nonlinearity $f$.
\begin{enumerate}
 \item[(fA)] The nonlinearity $f$ satisfies the
 estimate $$|f^{'}(\xi)|\leq c(1+|\xi|^{\alpha-1})$$ for some $\alpha\in
 [0,\frac{2}{(d-2)_{+}})$ (here $s_{+}=s$ if $s>0$ and $=0$ if $s\leq
 0$) and $$|f(\xi)|\leq c(1+|\xi|^{\beta})$$ for some $\beta\in[0,\frac{2}{d}).$
\end{enumerate}

The following theorem is proved in ~\cite{Oh3,Cazenave}.\\
\begin{them}
Assume that the nonlinearity $f$ satisfies the condition (fA), and
that the potential $V$ is bounded. Then Equation (~\ref{NLS}) is
globally well posed in $\mathcal{H}^{1}$, i.e. the Cauchy problem
for Equation (~\ref{NLS}) with a datum $\psi(0)\in \mathcal{H}^{1}$
has a unique solution $\psi(t)$ in the space $\mathcal{H}^{1}$ and
this solution depends continuously on $\psi(0)$. Moreover $\psi(t)$
satisfies the conservation laws (CE) and (CP).
\end{them}
\section{Existence and Orbital Stability of Solitons}\label{ExSt}
In this section we review the question of existence of the solitons
(ground states) for Equation (~\ref{NLS}). Assume the nonlinearity
$f:\mathbb{R}\rightarrow \mathbb{R}$ is smooth and satisfies
\begin{enumerate}
 \item[(fB)] There exists an interval $\mathcal{I}_{0}\in \mathbb{R}^{+}$
s.t. for any $\lambda\in\mathcal{I}_{0}$, $-\infty\leq
\displaystyle\overline{\lim}_{s\rightarrow +
 \infty}\frac{f(s)}{s^{\frac{2}{d-2}}}\leq 0$
and $\frac{1}{\xi}\int_{0}^{\xi}f(s)d
 s>\lambda$ for some constant $\xi$, for $d>2$;
and
 $$U(\phi,\lambda):=-\lambda\phi^{2}+\int_{0}^{\phi^{2}}f(\xi)d\xi$$
 has a smallest positive root $\phi_{0}(\lambda)$ such that
 $U_{\phi}(\phi_{0}(\lambda),\lambda)\not=0$, for $d=1$.
 \end{enumerate}

It is shown in ~\cite{BL, Strauss} that under Condition (fB) there
exists a spherical symmetric positive solution $\phi^{\lambda}$ to
the equation
\begin{equation}\label{soliton}
-\Delta\phi^{\lambda}+\lambda\phi^{\lambda}-f((\phi^{\lambda})^{2})\phi^{\lambda}=0.
\end{equation}
\begin{remark}
Existence of soliton functions $\phi^{\lambda}$ for $d=2$ is proved
in ~\cite{Strauss} under different conditions on $f$.
\end{remark}
When the potential $V$ is present, then some of the solitons above
bifurcate into solitons for Equation (~\ref{NLS}). Namely, let, in
addition, $f$ satisfy the condition $|f^{'}(\xi)|\leq
c(1+|\xi|^{p}),$ for some $p< \infty$, and $V$ satisfy the condition
\begin{enumerate}
\item[(VA)] $V$ is smooth and $0$ is a non-degenerate local
minimum of $V$.
\end{enumerate}
Then, similarly as in ~\cite{Floer,Oh1, ABC} one can show that if
$h$ is sufficiently small, then for any $\lambda\in
\mathcal{I}_{0V}$, where
$$\mathcal{I}_{0V}:=\{\lambda|\lambda>-\displaystyle\inf_{x\in\mathbb{R}}\{V(x)\}\}\cap\{\lambda|\lambda+V(0)\in
\mathcal{I}_{0}\},$$ there exists a unique soliton
$\phi^{\lambda}\equiv\phi_{h}^{\lambda}$ (i. e. $\phi^{\lambda}\in
\mathcal{H}_{2}(\mathbb{R})$ and $\phi^{\lambda}>0$) satisfying the
equation
$$-\Delta \phi^{\lambda}+(\lambda+V_{h})\phi^{\lambda}-
f((\phi^{\lambda})^{2})\phi^{\lambda}=0$$ and the estimate
$\phi^{\lambda}=\phi^{\lambda+V(0)}_{0}+O(h^{3/2})$ where
$\phi_{0}^{\lambda}$ is the soliton of Equation (~\ref{soliton}).

Let $\delta(\lambda):=\|\phi^{\lambda}\|_{2}^{2}$. It is shown in
~\cite{GSS1} that the soliton $\phi^{\lambda}$ is a minimizer of the
energy functional $H(\psi)$ for a fixed number of particles
$N(\psi)=constant$ if and only if
\begin{equation}\label{Stab}
\delta^{'}(\lambda)>0.
\end{equation}
Moreover, it shown in ~\cite{We2,GSS1} that under the latter
condition the solitary wave $\phi^{\lambda}e^{i\lambda t}$ is
orbitally stable.
Under more restrictive conditions (see ~\cite{GSS1}) on $f$ one can
show that the open set
%
\begin{equation}
\mathcal{I}:=\{\lambda\in \mathcal{I}_{0V}:\delta'(\lambda)>0\}
\end{equation}
is non-empty. Instead of formulating these conditions we assume in
what follows that the open set $\mathcal{I}$ is non-empty and
$\lambda\in \mathcal{I}$.

Using the equation for $\phi^{\lambda}$ one can show that if the
potential $V$ is redially symmetric then there exist constants $c,\
\delta>0$ such that
\begin{equation}\label{expondecay}
|\phi^{\lambda}(x)|\leq ce^{-\delta|x|}\ \text{and}\
|\frac{d}{d\lambda}\phi^{\lambda}|\leq ce^{-\delta|x|},
\end{equation}
and similarly for the derivatives of $\phi^{\lambda}$ and
$\frac{d}{d\lambda}\phi^{\lambda}$.

\section{Linearized Equation and Resonances}\label{subslinear} We
rewrite Equation (~\ref{NLS}) as $\frac{d\psi}{dt}=G(\psi)$ where
the nonlinear map $G(\psi)$ is defined by
\begin{equation}\label{symmetryG}
G(\psi)=-i(-\Delta+\lambda+V_{h})\psi+if(|\psi|^{2})\psi.
\end{equation}
Then the linearization of Equation (~\ref{NLS}) can be written as
$\frac{\partial\chi}{\partial t}=\partial G(\phi^{\lambda})\chi$
where $\partial G(\phi^{\lambda})$ is the Fr\'echet derivative of
$G(\psi)$ at $\phi$. It is computed to be
\begin{equation}\label{defineoperator}
\partial
G(\phi^{\lambda})\chi=-i(-\Delta+\lambda+V_{h})\chi+if((\phi^{\lambda})^{2})\chi+2if^{'}((\phi^{\lambda})^{2})(\phi^{\lambda})^{2}Re\chi.
\end{equation}
This is a real linear but not complex linear operator. To convert it
to a linear operator we pass from complex functions to real
vector-functions
$$\chi\longleftrightarrow \vec{\chi}=\left(
\begin{array}{lll}
\chi_{1}\\
\chi_{2}
\end{array}
\right),$$ where $\chi_{1}=Re\chi$ and $\chi_{2}=Im\chi.$ Then
$\partial G(\phi^{\lambda})\chi\longleftrightarrow
L(\lambda)\vec{\chi}$ where the operator $L(\lambda)$ is given by
\begin{equation}\label{operaL}
L(\lambda) :=  \left(
\begin{array}{lll}
0&L_{-}(\lambda)\\
-L_{+}(\lambda)&0
\end{array}
\right),
\end{equation}
with
\begin{equation}\label{firstoperator}
L_{-}(\lambda):=-\Delta+V_{h}+\lambda-f((\phi^{\lambda})^{2}),
\end{equation} and
\begin{equation}\label{secondoperator}
L_{+}(\lambda):=-\Delta+V_{h}+\lambda-f((\phi^{\lambda})^{2})-2f^{'}((\phi^{\lambda})^{2})(\phi^{\lambda})^{2}.
\end{equation}
Then we extend the operator $L(\lambda)$ to the complex space
$\mathcal{H}^{2}(\mathbb{R},\mathbb{C})\oplus
\mathcal{H}^{2}(\mathbb{R},\mathbb{C}).$

By a general result (see e.g. ~\cite{HS,RSIV}),
$$\sigma_{ess}(L(\lambda))=(-i\infty,-i\lambda]\cap
[i\lambda,i\infty)$$ if the potential $V_{h}$ in Equation
(~\ref{NLS}) decays at $\infty.$

The eigenfunctions of $L(\lambda)$ are described in the following
theorem (cf ~\cite{GS1}).

\begin{theorem}\label{mainpo}
Let $V$ satisfy Condition (VA). Then the operator $L(\lambda)$ has
at least $2d +2$ eigenvectors and associated eigenvectors with
eigenvalues near near zero: two-dimensional space with the
eigenvalue 0 and a $2d $-dimensional space with non-zero imaginary
eigenvalues.
\end{theorem}
\begin{proof}
We already know that the vector $\left(
\begin{array}{lll}
0\\
\phi^{\lambda}
\end{array}
\right)$ is an eigenvector of $L(\lambda)$ with eigenvalue $0$ and
$\left(
\begin{array}{lll}
\partial_{\lambda}\phi^{\lambda}\\
0
\end{array}
\right)$ is an associated eigenvector,
\begin{equation}\label{eigenvalue0}
L(\lambda)\left(
\begin{array}{lll}
0\\
\phi^{\lambda}
\end{array}
\right)=0,\ L(\lambda)\left(
\begin{array}{lll}
\partial_{\lambda}\phi^{\lambda}\\
0
\end{array}
\right)=\left(
\begin{array}{lll}
0\\
\phi^{\lambda}
\end{array}
\right).
\end{equation}

Similarly as in ~\cite{GS1} one can show that the operator
$L(\lambda)$ has also the eigenvalues $\pm i\epsilon_j(\lambda),$
$\epsilon_j(\lambda)>0$, with the eigenfunctions
$\left(\begin{array}{lll}
\xi_j\\
\pm i\eta_j
\end{array}
\right)$, related by complex conjugation. Moreover,
$\epsilon_j(\lambda):=h\sqrt{2e_j} +o(h)$ where $e_j$ are
eigenvalues of the Hessian matrix of $V$ at value $x=0,$
$V^{''}(0)$, and
$$
\xi_j= \sqrt{2}\displaystyle \partial_{x_{k}}\phi_{0}^{\lambda}
 +o(h)\
\text{and}\ \eta_j= -h \sqrt{e_j}\displaystyle
x_{j}\phi^{\lambda}_{0} +o(h),$$ and $\xi_{i}$ and $\eta_{j}$ are
real.
\end{proof}
\begin{remark}
The zero eigenvector $\left(
\begin{array}{lll}
0\\
\phi^{\lambda}
\end{array}
\right)$ and the associated zero eigenvector $\left(
\begin{array}{lll}
\partial_{\lambda}\phi^{\lambda}\\
0
\end{array}
\right)$ are related to the gauge symmetry $\psi(x,t)\rightarrow
e^{i\alpha}\psi(x,t)$ of the original equation and the $2d$
eigenvectors $\left(\begin{array}{lll}
\xi_j\\
\pm i\eta_j
\end{array}
\right)$ with $O(h)$ eigenvalues originate from the zero
eigenvectors $\left(
\begin{array}{lll}
\partial_{x_{k}}\phi_{0}^{\lambda}\\
0
\end{array}
\right), k=1,2,\cdot\cdot\cdot,d,$ and the associated zero
eigenvectors $\left(
\begin{array}{lll}
0\\
x_{k}\phi_{0}^{\lambda}
\end{array}
\right),\ k=1,2,\cdot\cdot\cdot,d,$ of the $V=0$ equation due to the
translational symmetry and to the boost transformation
$\psi(x,t)\rightarrow e^{ib\cdot x}\psi(x,t)$ (coming from the
Galilean symmetry), respectively.
\end{remark}

We say that a function $g\in \mathcal{L}^{2}(\mathbb{R}^{d})$ is
permutational symmetric if
$$g(x)=g(\sigma x)\ \text{for any}\ \sigma\in S_{d}$$ with $S_{d}$ being the group of
permutation of $d$ indices and
$$\sigma
(x_{1},x_{2},\cdot\cdot\cdot,x_{d}):=(x_{\sigma(1)},x_{\sigma(2)},\cdot\cdot\cdot,x_{\sigma(d)}).$$
\begin{remark}\label{remark2}
For any function of the form $e^{ip\cdot x}\phi(|x-a|)$ with $a
\parallel p$, there exists a rotation $\tau$ such that
the function $e^{ip\cdot \tau
{x}}\phi(|\tau{x}-a|)=e^{i\tau^{-1}p\cdot x}\phi(|x-\tau^{-1}a|)$ is
permutational symmetric. Such families describe wave packets with
the momenta directed toward or away from the origin.
\end{remark} If for $d\geq 2$ the potential $V(x)$ is spherically
symmetric, then $V^{''}(0)=\frac{1}{d}\Delta V(0)\cdot Id_{n\times
n}$, and therefore the eigenvalues $e_{j}$ of $V^{''}(0)$ are all
equal to $ \frac{1}{d}\Delta V(0)$. Thus we have
\begin{corollary}\label{mainpo1}
Let $d\geq 2$ and $V$ satisfy Condition (VA) and let $V$ be
spherically symmetric. Then $L(\lambda)$ restricted to permutational
symmetric functions has $4$ eigenvectors or associated eigenvectors
near zero: two-dimensional space with eigenvalue 0; and
two-dimensional space with the non-zero imaginary eigenvalues $\pm
i\epsilon(\lambda)$, where $\epsilon(\lambda)=h\sqrt{\frac{2\Delta
V(0)}{d}}+o(h)$, and with the eigenfunctions
$\left(\begin{array}{lll}
\xi\\
\pm i\eta
\end{array}
\right)$, where $\xi$ and $\eta$ are real, and permutation symmetric
functions satisfying
$$\xi=\sqrt{2}\displaystyle\sum_{n=1}^{d}\frac{d}{dx_{n}}\phi_{0}^{\lambda}
+O(h)\ \text{and}\ \eta=-h\sqrt{\frac{1}{d}\Delta
V(0)}\displaystyle\sum_{n=1}^{d}x_{n}\phi^{\lambda}_{0} +O(h^{3/2})
.$$
\end{corollary}

The eigenvectors $\left(
\begin{array}{lll}
\xi\\
\pm i\eta
\end{array}
\right)$ are symmetric combinations of the eigenvectors described in
the proof of Theorem 2. Observe that
\begin{equation}\label{SymNo}
Span\{\phi^{\lambda},\phi^{\lambda}_{\lambda}\}\perp
Span\{\xi,\eta\}\end{equation} since
$\phi^{\lambda},\phi^{\lambda}_{\lambda}$ are spherically symmetric,
and
\begin{equation}\label{xieta}
\langle \xi,\eta\rangle =\frac{1}{\epsilon(\lambda)}\langle
L_{-}(\lambda)\eta,\eta\rangle>0.
\end{equation}

Besides eigenvalues, the operator $L(\lambda)$ may have resonances
at the tips, $\pm i\lambda$, of its essential spectrum (those tips
are called thresholds). To define the resonance we write the
operator $L(\lambda)$ as
$L(\lambda)=L_{0}(\lambda)+V_{big}(\lambda),$ where
\begin{equation}\label{Lo}
L_{0}(\lambda):=\left(
\begin{array}{lll}
0&-\Delta+\lambda\\
\Delta-\lambda&0
\end{array}
\right),
\end{equation}
and
\begin{equation}\label{Vbig}
V_{big}(\lambda):=\left(
\begin{array}{lll}
0&V_{h}-f((\phi^{\lambda})^{2})\\
-V_{h}+f((\phi^{\lambda})^{2})+2f^{'}((\phi^{\lambda})^{2})(\phi^{\lambda})^{2}&0
\end{array}
\right).
\end{equation}

Recall the notation $\alpha_+ := \alpha$ if $\alpha>0$ and $=0$ of
$\alpha\leq 0$.
\begin{definition}
Let $d\neq 2$. A function $h$ is called a resonance function of
$L(\lambda)$ at $\mu=\pm i\lambda$ if $h\not\in \mathcal{L}^{2}$,
$|h(x)|\leq c\langle x\rangle^{-(d-2)_+}$ and $h$ is $C^{2}$ and
solves the equation
$$(L(\lambda)-\mu)h=0.$$
\end{definition}

Note that this definition implies that for $d>2$ the resonance
function $h$ solves the equation
$$(1+K(\lambda))h=0$$
where $K(\lambda)$ is a family of compact operators given by
$K(\lambda) := (L_{0}(\lambda)-\mu+0)^{-1}V_{big}(\lambda)$.

In this paper we make the following assumptions for the point
spectrum and resonances of the operator $L(\lambda):$
\begin{enumerate}
 \item[(SA)] $L(\lambda)$ has only $4$ standard and associated
eigenvectors in the permutation symmetric subspace.
 \item[(SB)] $L(\lambda)$ has no resonances at $\pm i\lambda$.
\end{enumerate}

The discussion and results concerning these conditions, given in
~\cite{GS1}, suggested strongly that Condition (SA) is satisfied for
a large class of nonlinearities and potentials and Condition (SB) is
satisfied generically. In ~\cite{GSV} we show this using earlier
results of ~\cite{CP, CPV}. We also assume the following condition
\begin{enumerate}
 \item[(FGR)] Let $N$ be the smallest positive integer
such that $\epsilon(\lambda)(N+1)> \lambda\ \forall \lambda \in I$.
Then $Re Y_{N}< 0$ where $Y_{n},\ n=1,2,\cdots,$ are the functions
of $V$ and $\lambda,$ defined in Lemma ~\ref{transformz} below.
\end{enumerate}

We expect that Condition (FGR) holds generically. Theorem
~\ref{maintheorem2} below shows that $Re Y_{n}= 0$ if $n<N.$

We expect the following is true: (a) if for some $N_{1}(\geq N),$
$ReY_{n}=0$ for $n<N_{1},$ then $ReY_{N_{1}}\leq 0$ and (b) for
generic potentials and nonlinearities there exists an $N_{1}(\geq
N)$ such that $ReY_{N_{1}}\not=0.$ Thus Condition (FGR) could have
been generalized by assuming that $Re Y_{N_1}< 0$ for some $N_1 \geq
N$ such that $ReY_{n}=0$ for $n<N_{1}$. We took $N=N_1$ in order not
to complicate the exposition.

The following form of $Re Y_{N}$
 \begin{equation}\label{FGR}
 Re Y_{N}=Im\langle \sigma_{1}(L(\lambda)-(N+1)i\epsilon(\lambda)-0)^{-1}F,F\rangle\leq 0
 \end{equation}
 for some function
 $F$ depending on $\lambda$ and $V$ and the matrix $\sigma_{1}:=\left(
 \begin{array}{lll}
 0&-1\\
 1&0
 \end{array}
 \right)$, is proved in ~\cite{BuSu, TY1, TY2, SW} for $N=1,$ and in ~\cite{G} for $N=2,3$.
 We conjecture that this formula holds for any $N$.

 Condition (FGR) is related to the Fermi Golden Rule condition which
 appears whenever time-(quasi)periodic, spatially localized solutions
 become coupled to radiation. In the standard case it says that this coupling is effective
 in the second order ($N=1$) of the perturbation theory and therefore it leads to instability of such
 solutions. In our case these time-periodic solutions are
 stationary solutions
$$c_1 \left(
\begin{array}{lll}
\xi\\
 i\eta
\end{array}
\right)e^{i\epsilon(\lambda) t} +c_2 \left(
\begin{array}{lll}
\xi\\
- i\eta
\end{array}
\right)e^{-i\epsilon(\lambda) t}$$
 of the linearized equation $\frac{\partial\vec{\chi}}{\partial
 t}=L(\lambda)\vec{\chi}$ and the coupling is realized through the
 nonlinearity. Since the radiation in our case is "massive"$-$ the
 essential spectrum of $L(\lambda)$ has the gap
 $(-i\lambda,i\lambda)$, $\lambda>0,$ $-$ the coupling occurs only in
 the $N-$th order of perturbation theory where $N$ is the same as in
 Condition (FGR).

The rigorous form of the Fermi Golden Rule for the linear
Schr\"odinger  equations was introduced in ~\cite{Simon} (see
~\cite{RSIV}). For nonlinear waves and Schr\"odinger equations the
Fermi Golden Rule and the corresponding condition were introduced in
~\cite{S} and, in the present context, in ~\cite{CLR, SW, BuSu, BP2,
TY1, TY2, TY3}.

\section{Main Results}\label{sectionmaintheorem}
In this section we state the main theorem of this paper. For
technical reason we impose the following conditions on $f$ and $V$
\begin{enumerate}
 \item[(fC)] the nonlinearity $f$ is a smooth function satisfying $f^{''}(0)=f^{'''}(0)=0$
 if $d\geq 3$;
and $f^{(k)}(0)=0$ for $k=2,3\cdot\cdot\cdot 3N+1$ if $d=1,$ where
$f^{(k)}$ is the $k-$th derivative of $f$, and $N$ is the same as in
Condition (FGR),
 \item[(VB)] $V$ decays exponentially fast at $\infty.$
\end{enumerate}
\begin{theorem}\label{maintheorem1}
Let Conditions (fA)-(fC), (VA), (VB), (SA), (SB) and (FGR) be
satisfied and let, for $d\geq 3$, the potential $V$ be spherically
symmetric. Let an initial condition $\psi_0$ be permutation
symmetric if $d\geq 3$ and $\lambda\in \mathcal{I}$. There exists
$c>0$ such that, if
\begin{equation}\label{InitCond}\inf_{\gamma\in
\mathbb{R}}\{\|\psi_0-e^{i\gamma}(\phi^{\lambda}+z_{1}^{(0)}\xi+iz_{2}^{0}\eta)\|_{\mathcal{H}^{k}}+
\|(1+x^{2})^{\nu}[\psi_0-e^{i\gamma}(\phi^{\lambda}+z_{1}^{(0)}\xi+iz_{2}^{0}\eta)]\|_{2}\}\leq
c[(z_{1}^{0})^{2}+(z_{2}^{0})^{2}]\end{equation} with small real
constants $z_{1}^{(0)}$ and $z_{2}^{(0)}$, some large constant
$\nu>0$ and with $k=[\frac{d}{2}]+3$ if $d\geq 3,$ and $k=1$ if
$d=1,$ then there exist differentiable functions $\gamma,\ z_{1},\
z_{2}:\mathbb{R}^{+}\rightarrow \mathbb{R},$
$\lambda:\mathbb{R}^{+}\rightarrow \mathcal{I}$ and
$R:\mathbb{R}^{+}\rightarrow \mathcal{H}^{k}$ such that the
solution, $\psi(t)$, to Equation (~\ref{NLS}) is of the form
\begin{equation}\label{Parametrization}
\psi(t)=e^{i \int_{0}^{t}
\lambda(s)ds+i\gamma(t)}[\phi^{\lambda(t)}+
z_{1}(t)\xi+iz_{2}(t)\eta+R(t)]
\end{equation} with the following
estimates:
\begin{enumerate}
 \item[(A)] $\displaystyle\|(1+x^{2})^{-\nu}R(t)\|_{2}\leq c(1+|t|)^{-\frac{1}{N}}$ where $\nu$ is the same as in (~\ref{InitCond}) and $N$ is the same as that in
 (FGR),
 \item[(B)] $\displaystyle\sum_{j=1}^{2}|z_j(t)|\leq c(1+t)^{-\frac{1}{2N}}.$
\end{enumerate}
\end{theorem}
\begin{remark}
Recall from Remark ~\ref{remark2} that the class of permutationally
symmetric data includes wave packets with initial momenta directed
toward or in the opposite direction of the origin.
\end{remark}
\begin{theorem}\label{maintheorem2} Under the conditions of Theorem
3 we have
\begin{enumerate}
 \item[(A)] there exists a constant $\lambda_{\infty}\in \mathcal{I}$ such
 that $\displaystyle\lim_{t\rightarrow \infty}\lambda(t)=\lambda_{\infty}.$
 \item[(B)] Let
 $z:=z_{1}-iz_{2}$. Then there exists a change of variables
 $\beta=z+O(|z|^{2})$ such that
 \begin{equation}\label{betadecay}
 \dot{\beta}=i\epsilon(\lambda)\beta+\sum_{n=1}^{N}Y_{n}(\lambda)\beta^{n+1}\bar{\beta}^{n}+O(|\beta|^{2N+2})
 \end{equation}
\end{enumerate}
 with $Y_{n}$ being purely imaginary if $n<N$ and, by Condition (FGR) $Re
 Y_{N}<
 0$. Moreover,
 for $N=1,2,3$,  $ReY_{N}$ is given by Equation (~\ref{FGR}).
(Recall that $\epsilon(\lambda)=h\sqrt{\frac{2\Delta
V(0)}{d}}+o(h)$.)
\end{theorem}
\begin{remark}
Equations (~\ref{Parametrization}) and (~\ref{betadecay}) can be
rewritten in the form (~\ref{decom1}) with $a(t)$ and $p(t)$
satisfying the equations $\frac{1}{2}\dot{a}=p$ and
$\dot{p}=-h^{2}\nabla V(a)$ modulo $O(|a|^{2}+|p|^{2}).$
\end{remark}
The proof of Theorems ~\ref{maintheorem1} and ~\ref{maintheorem2}
are given in Sections ~\ref{secasymptotic}-~\ref{ProveMain} for
$d\geq 3$ and in Section ~\ref{Dimen1} for $d=1.$ In order not to
clutter the notation we restrict the arguments in Section
~\ref{ProveMain} to the case $d=3$ only.
\section{Re-parametrization of $\psi(t)$}\label{secasymptotic}
In this section we introduce a convenient decomposition of the
solution $\psi(t)$ to Equation (~\ref{NLS}) into a solitonic
component and a simplectically orthogonal fluctuation.
\begin{theorem}
There exists a constant $\delta>0$ such that if an initial condition
$\psi(0)$ satisfies $\displaystyle\inf_{\gamma\in
[0,2\pi)}\|\psi(0)-e^{i\gamma}\phi^{\lambda}\|_{\mathcal{H}^{1}}<\delta,$
then for any time $t$ $\psi(t)$ can be decomposed uniquely as
\begin{equation}\label{decompositionR}
\psi(t)=e^{i \int_{0}^{t}
\lambda(s)ds+i\gamma(t)}(\phi^{\lambda}+z_{1}(t)\xi+iz_{2}(t)\eta+R(t)),
\end{equation}
where $\lambda,\ \gamma,\ z_1,\ z_2 $ are real differentiable
functions of $t$, and the remainder $R(t)$ satisfies the
orthogonality conditions
\begin{equation}\label{Rorthogonal}
Im\langle R,i\phi^{\lambda}\rangle=Im\langle
R,\frac{d}{d\lambda}\phi^{\lambda}\rangle=Im\langle
R,i\eta\rangle=Im\langle R,\xi\rangle=0.
\end{equation}
\end{theorem}
\begin{proof}
By the Lyapunov stability (see ~\cite{GSS1}), $\forall\ \epsilon>0,$
there exists a constant $\delta,$ such that if
$\displaystyle\inf_{\gamma\in
R}\|\psi(0)-e^{i\gamma}\phi^{\lambda}\|_{\mathcal{H}^{1}}<\delta,$
then $\forall \ t>0$, $\displaystyle\inf_{\gamma}
\|\psi(t)-e^{i\gamma}\phi^{\lambda}\|_{\mathcal{H}^{1}}<\epsilon.$
Then Decomposition (~\ref{decompositionR}) (~\ref{Rorthogonal})
follows from Splitting Theorem in ~\cite{FGJS}.
\end{proof}
After plugging Equation (~\ref{decompositionR}) into Equation
(~\ref{NLS}), we get the equation
\begin{equation}\label{eq:R1R2}
\begin{array}{lll}
iR_{t}&=&\mathcal{L}(\lambda)R+N(R,z_{1},z_{2})+\epsilon(\lambda)[iz_{2}\xi+z_{1}\eta]
+\dot{\gamma}[\phi^{\lambda}+z_{1}\xi+iz_{2}\eta+R]\\
&
&-i\dot\lambda\phi^{\lambda}_{\lambda}-i\dot{z}_{1}\xi-i\dot{\lambda}z_{1}\partial_{\lambda}\xi+\dot{z}_{2}\eta+\dot\lambda
z_{2}\partial_{\lambda}\eta.
\end{array}
\end{equation}
where $\mathcal{L}(\lambda)$ is a real-linear operator given by
$$\mathcal{L}(\lambda)R:=-\Delta R+\lambda
R+V_{h}R-f((\phi^{\lambda})^{2})R-2f^{'}((\phi^{\lambda})^{2})(\phi^{\lambda})^{2}ReR,$$
and $N(R,z_{1},z_{2})$ is the nonlinear term given by
\begin{equation}\label{nonlinear}
\begin{array}{lll}
N(R, z_{1},z_{2})&:=&-f(|\phi^{\lambda}+z_{1}\xi+iz_{2}\eta+R|^{2})(\phi^{\lambda}+z_{1}\xi+iz_{2}\eta+R)\\
&
&+f((\phi^{\lambda})^{2})(\phi^{\lambda}+z_{1}\xi+iz_{2}\eta+R)+2f^{'}((\phi^{\lambda})^{2})(\phi^{\lambda})^{2}[z_{1}\xi+ReR].
\end{array}
\end{equation}

Projecting Equation (~\ref{eq:R1R2}) onto the vectors
$\phi^{\lambda},\ \phi^{\lambda}_{\lambda},\ \eta$ and $\xi$ we
derive the following equations for $\lambda,\ \gamma,\ z_{1}$ and
$z_{2}$ as follows
\begin{equation}\label{EqLamBeta}
\begin{array}{lll}
& &\dot\lambda[\delta^{'}(\lambda)-Re\langle
R,\phi^{\lambda}_{\lambda}\rangle]-\dot\gamma Im\langle
R,\phi^{\lambda}\rangle=Im\langle
N(R,z),\phi^{\lambda}\rangle,\\
& &\dot\gamma[\delta^{'}(\lambda)+Re\langle
R,\phi^{\lambda}_{\lambda}\rangle]-\dot\lambda Im\langle
R,\phi^{\lambda}_{\lambda\lambda}\rangle=-Re\langle
N(R,z),\phi^{\lambda}_{\lambda}\rangle,\\
\end{array}
\end{equation}
and
\begin{equation}\label{eq:z1z2}
\begin{array}{lll}
& &[\dot{z}_{1}-\epsilon(\lambda)z_{2}]\langle \xi,\eta\rangle\\
&=&\dot\lambda Re\langle R,\eta_{\lambda}\rangle+Im\langle
N(\vec{R},z),\eta\rangle+\dot\gamma z_{2}\langle
\eta,\eta\rangle+\dot\gamma Im\langle R,\eta\rangle-\dot\lambda
z_{1}\langle
\xi_{\lambda},\eta\rangle;\\
& &[\dot{z}_{2}+\epsilon(\lambda)z_{1}]\langle \xi,\eta\rangle\\
&=&\dot\lambda Im\langle R,\xi_{\lambda}\rangle-Re \langle
N(\vec{R},z),\xi\rangle-\dot\gamma z_{1}\langle
\xi,\xi\rangle-\dot\gamma Re\langle R,\xi\rangle-\dot\lambda
z_{2}\langle \eta_{\lambda},\xi\rangle.
\end{array}
\end{equation}

As was already discussed above since the operator
$\mathcal{L}(\lambda)$ is only real-linear we pass from the unknown
$R$ to the unknown $\vec{R}:=\left(
\begin{array}{lll}
Re R\\
Im R
\end{array}
\right)\leftrightarrow R$. Under this correspondence the
multiplication by $i^{-1}$ goes over to the symplectic matrix
$$J:=\left(
\begin{array}{lll}
0&1\\
-1&0
\end{array}
\right):\ J\vec{R}\leftrightarrow i^{-1}R.$$ Observe that due to
(~\ref{Rorthogonal})
\begin{equation}
\vec{R}\perp J\left(
\begin{array}{lll}
0\\
\phi^{\lambda}
\end{array}
\right), \ J\left(
\begin{array}{lll}
\phi_{\lambda}^{\lambda}\\
0
\end{array}
\right),\ J\left(
\begin{array}{lll}
\xi\\
0
\end{array}
\right),\ J\left(
\begin{array}{lll}
0\\
\eta
\end{array}
\right).\end{equation} On the other hand in Equations
(~\ref{eq:z1z2}) it is more convenient to go from the real,
symplectic structure given by $J$ to the complex structure
$i^{-1}$ by passing from $\left(
\begin{array}{lll}
z_{1}\\
z_{2}
\end{array}
\right)$ to $z:=z_{1}-iz_{2}$. Let $\vec{N}(\vec{R},z):= \left(
\begin{array}{lll}
Re N(R,z_{1},z_{2})\\
Im N(R,z_{1},z_{2})
\end{array}
\right).$ Then
\begin{equation}\label{eq:vectorform}
\frac{d}{dt}\vec{R}=L(\lambda)\vec{R}+\dot\gamma J
\vec{R}+J\vec{N}(\vec{R},z)+\left(
\begin{array}{lll}
z_{2}\epsilon(\lambda)\xi+\dot\gamma
z_{2}\eta-\dot\lambda\phi^{\lambda}_{\lambda}-\dot{z}_{1}\xi-\dot{\lambda}z_{1}\xi_{\lambda}\\
-z_{1}\epsilon(\lambda)\eta-\dot{\gamma}\phi^{\lambda}-\dot{\gamma}z_{1}\xi-\dot{z}_{2}\eta-\dot{\lambda}z_{2}\eta_{\lambda}
\end{array}
\right)
\end{equation} where $z_1=$ Re$z, z_2=$ Im$z$ and the linear operator $L(\lambda)$ is
given by (~\ref{operaL})-(~\ref{secondoperator}).

Define $P_{d}$ as the Riez projection for the isolated eigenvalues
of $L(\lambda)$. It was shown in ~\cite{GS1} that (in the Dirac
notation)
\begin{equation}\label{eq:concreteform}
\begin{array}{lll}
P_{d}&=&\frac{1}{\delta^{'}(\lambda)}(\left|
\begin{array}{lll}
0\\
\phi^{\lambda}
\end{array}
\right\rangle \left\langle
\begin{array}{lll}
\frac{d}{d\lambda}\phi^{\lambda}\\
0
\end{array}
\right|+\left|
\begin{array}{lll}
\frac{d}{d\lambda}\phi^{\lambda}\\
0
\end{array}
\right\rangle \left\langle
\begin{array}{lll}
0\\
\phi^{\lambda}
\end{array}
\right|)\\
& &+\frac{i}{2\langle \xi,\eta\rangle}(\left|
\begin{array}{lll}
\xi\\
i\eta
\end{array}
\right\rangle\left\langle
\begin{array}{lll}
-i\eta\\
\xi
\end{array}
\right| + \left|
\begin{array}{lll}
-\xi\\
i\eta
\end{array}
\right\rangle\left\langle
\begin{array}{lll}
i\eta\\
\xi
\end{array}
\right|).
\end{array}
\end{equation} We denote $P_{c}:=1-P_{d}.$ We call
$P_{c}$ the projection onto the essential spectrum of $L(\lambda)$.

Since $P_{c}\vec{R}=\vec{R}$, we have that
$$P_{c}\frac{d}{dt}\vec{R}=\frac{d}{dt}\vec{R}-\dot\lambda P_{c\lambda}\vec{R}.$$
Applying the projection $P_{c}$ to Equation (~\ref{eq:vectorform})
and using the relations above we find
\begin{equation}\label{eq:R1R22}
\begin{array}{lll}
\frac{d}{dt}\vec{R}&=&L(\lambda)\vec{R}+\dot\lambda
P_{c\lambda}\vec{R}+\dot\gamma P_{c}
J\vec{R}+P_{c}J\vec{N}(\vec{R},z)\\&+&\frac{1}{2}\dot\gamma
P_{c}[z\left(
\begin{array}{lll}
-i\eta\\
\xi
\end{array}
\right)+\bar{z}\left(
\begin{array}{lll}
i\eta\\
\xi
\end{array}
\right)]-\frac{1}{2}\dot\lambda P_{c}[z\left(
\begin{array}{lll}
\xi_{\lambda}\\
-i\eta_{\lambda}
\end{array}
\right)+\bar{z}\left(
\begin{array}{lll}
\xi_{\lambda}\\
i\eta_{\lambda}
\end{array}
\right)].
\end{array}
\end{equation}

On the other hand Equations (~\ref{eq:z1z2}) for $z_{1}$ and $z_{2}$
become
\begin{equation}\label{eq:z2}
\begin{array}{lll}
\dot{z}&=&i\epsilon(\lambda)z+\frac{1}{\langle
\xi,\eta\rangle}\langle J\vec{N}(\vec{R},z)+\dot\gamma\left(
\begin{array}{lll}
z_{2}\eta\\
z_{1}\xi
\end{array}
\right)+\dot\gamma J \vec{R}-\dot\lambda\left(
\begin{array}{lll}
z_{1}\xi_{\lambda}\\
-z_{2}\eta_{\lambda}
\end{array}
\right),\left(
\begin{array}{lll}
\eta\\
-i\xi
\end{array}
\right)\rangle\\
& &+\frac{\dot{\lambda}}{\langle \xi,\eta\rangle}\langle
\vec{R},\left(
\begin{array}{lll}
\eta_{\lambda}\\
-i\xi_{\lambda}
\end{array}
\right)\rangle.
\end{array}
\end{equation}
Finally, Equation (~\ref{EqLamBeta}) for $\lambda$ and $\gamma$ can
be rewritten as
\begin{equation}\label{eq:lambda}
\left(
\begin{array}{lll}
\delta^{'}(\lambda)+\langle
R_{1},\phi_{\lambda}^{\lambda}\rangle&-\langle
R_{2},\phi_{\lambda\lambda}^{\lambda}\rangle\\
-\langle R_{2},\phi^{\lambda}\rangle&\delta^{'}(\lambda)-\langle
R_{1},\phi_{\lambda\lambda}^{\lambda}\rangle
\end{array}
\right)\left(
\begin{array}{lll}
\dot\gamma\\
\dot\lambda
\end{array}
\right)=\left(
\begin{array}{lll}
-Re\langle N(\vec{R},z),\phi_{\lambda}^{\lambda}\rangle\\
Im\langle N(\vec{R},z),\phi^{\lambda}\rangle
\end{array}
\right).
\end{equation}
\begin{remark}
By the gauge invariance of Equation (~\ref{NLS}), Equations
(~\ref{eq:R1R22})-(~\ref{eq:lambda}) are invariant under the gauge
transformation, $\gamma\rightarrow \gamma+\alpha,$ for any
$\alpha\in \mathbb{R}$, and other parameters fixed. Hence these
equations and their solutions are independent of $\gamma.$
\end{remark}
\section{Expansions of the Functions $\vec{R},\  \dot\lambda\ \text{and}\
\dot\gamma$}\label{Sec:expan} In this section we construct
expansions of the functions $\vec{R},\ \dot\lambda,\ \dot{z}\
\text{and}\ \dot\gamma$ in the parameter
$$z:=z_{1}-iz_{2}.$$ In what follows we fix $N$ to be the smallest positive integer such
that $(N+1)\epsilon(\lambda)>\lambda,$ where, recall, that
$i\epsilon(\lambda)$ and $-i\epsilon(\lambda)$ are the only nonzero
eigenvalues of $L(\lambda).$
\begin{definition}\label{admissible}
A vector-function $\vec{u}:\ \mathbb{R}^{d}\rightarrow
\mathbb{C}^{2}$ is admissible if the vector-function $\left(
\begin{array}{lll}
1&0\\
0&i
\end{array}
\right)\vec{u}$ has real entries.
\end{definition}
\begin{theorem}\label{expansion}
There exists a polynomial $P(z,\bar{z})=\displaystyle\sum_{2\leq
m+n\leq N}a_{m,n}(\lambda)z^{m}\bar{z}^{n}$ with
$a_{m,n}(\lambda)\in \mathbb{R}$ such that if we define
$y:=z+P(z,\bar{z})$ then for any $2\leq k\leq 2N,$ the function
$\vec{R}$ can be decomposed as
\begin{equation}\label{expansionR1}
\vec{R}=\displaystyle\sum_{2\leq m+n\leq
k}R_{mn}(\lambda)y^{m}\bar{y}^{n}+R_{k}
\end{equation}
where the functions $R_{mn}(\lambda),\
R_{k}:\mathbb{R}^{3}\rightarrow \mathbb{C}^{2}$ have the following
properties:
\begin{enumerate}
 \item[(RA)] if $\max\{m,n\}\leq N,$ then the functions $R_{m,n}(\lambda)\in \mathcal{L}^{2}$ are admissible, and decay
 exponentially fast at $\infty;$
 \item[(RB)] if $\max\{m,n\}>N$ then
 the functions $R_{m,n}(\lambda)$ are of the form
\begin{equation}\label{eq:unusual}
\prod_{k}(L(\lambda)-ik\epsilon(\lambda)+0)^{-n_{k}}P_{c}\phi_{m,n}(\lambda),
\end{equation}
where the functions $\phi_{m,n}(\lambda)$ are smooth and decay
exponentially fast at $\infty,$ $0\leq
\displaystyle\sum_{k}n_{k}\leq N,$ and $2N\geq k\geq N+1;$ note that
the equation (~\ref{eq:unusual}) makes sense in an appropriate
weighted $\mathcal{L}^{2}$ space (see Section
~\ref{Sec:Propagator});
\item[(RC)]
the function $R_{k}$ $(N\leq k\leq 2N)$ satisfies the equation
\begin{equation}\label{equationg}
\begin{array}{lll}
\frac{d}{dt}R_{k}&=&L(\lambda)R_{k}+P_{k}(y,\bar{y})R_{k}+N_{N}(R_{N},y,\bar{y})+F_{k}(y,\bar{y}),
\end{array}
\end{equation} where
\begin{enumerate}
 \item[(1)] $F_{k}(y,\bar{y})=O(|y|^{k+1})$ is a polynomial in $y$ and
 $\bar{y}$ with $\lambda$-function-valued coefficients, and each coefficient can be written as the sum of
 functions of the form (~\ref{eq:unusual});
 \item[(2)] $P_{k}(y,\bar{y})$ is the operator defined by
 $$P_{k}(y,\bar{y}):=\dot{\gamma}P_{c}J+\dot{\lambda}P_{c\lambda}+A_{k}(y,\bar{y}),$$
 where $A_{k}(y,\bar{y})$ is a $2\times 2$ matrix-valued function of $y,\ \bar{y},\ x$ and $\lambda$ bounded in the matrix norm as $$|A_{k}(y,\bar{y})|\leq c|y|e^{-\epsilon_{0}|x|};$$
 \item[(3)] $N_{N}(R_{N},y,\bar{y})$
 satisfies the estimates
 \begin{equation}\label{remainderestimate}
 \|N_{N}(R_{N},y,\bar{y})\|_{1}+\|N_{N}(R_{N},y,\bar{y})\|_{\mathcal{H}^{l}}\leq c
 (1+t)^{-\frac{2N+3}{2N}}[Y^{2}\mathcal{R}_{1}^{6}+\mathcal{R}_{1}^{5}\mathcal{R}_{2}^{2}],
 \end{equation}
 \begin{equation}\label{remainderestiamte2}
 \begin{array}{lll}
 &
 &\|(-\Delta+1)^{\frac{l}{2}}N_{N}(R_{N},y,\bar{y})\|_{1}+\|N_{N}(R_{N},y,\bar{y})\|_{\mathcal{H}^{l}}\\
 &\leq&c(T_{0}+t)^{-\frac{2N+3}{2N}}(Y^{2}\mathcal{R}_{1}^{2}\mathcal{R}_{2}^{2}+Y^{2}\mathcal{R}_{1}^{3}\mathcal{R}_{2}^{3}+\mathcal{R}_{1}^{3}\mathcal{R}_{2}^{4}),
 \end{array}
 \end{equation}
where the constant $l$ is defined as $l:=[\frac{d}{2}]+3,$ the
estimating functions $Y,\ \mathcal{R}_{1}$ and $\mathcal{R}_{2}$
depend on $t$ and are defined in Equations (~\ref{majorant}) below.
 \end{enumerate}
\end{enumerate}
 Furthermore, for $z$ satisfying Equation
 (~\ref{eq:z2}), the parameter $y$ satisfies the equation
 \begin{equation}\label{expanz2}
\dot{y}=i\epsilon(\lambda)y+\sum_{2\leq m+n\leq
2N+1}\Theta_{mn}(\lambda)y^{m}\bar{y}^{n}+Remainder,
\end{equation}
where $\Theta_{mn}(\lambda)$ is purely imaginary for $m,n\leq N$;
$\Theta_{m,n}(\lambda)=0$ for $m+n\leq N$ and $m\not=n+1$. The term
$Remainder$ is bounded as
\begin{equation}\label{remainder}
|Remainder(t)|\leq
c(T_{0}+t)^{-\frac{2N+2}{2N}}(Y(t)+\mathcal{R}_{1}(t))^{2}.
\end{equation}
\end{theorem}
Above we used the following functions:
\begin{equation}\label{majorant}
\begin{array}{lll}
Y(T):=\displaystyle\max_{t\leq T}(T_{0}+t)^{\frac{1}{2N}}|y(t)|,\\
\mathcal{R}_{1}(T):=\displaystyle\max_{t\leq
T}[(T_{0}+t)^{\frac{N+1}{2N}}(\|\rho_{\nu}R_{N}\|_{\mathcal{H}^{l}}+\|R_{N}(t)\|_{\infty})+(T_{0}+t)^{\frac{2N+1}{2N}}\|\rho_{\nu}R_{2N}(t)\|_{2}]\\
\mathcal{R}_{2}(T):=\displaystyle\max_{t\leq
T}\|R_{N}(t)\|_{\mathcal{H}^{l}}
\end{array}
\end{equation}
where $l:=[\frac{d}{2}]+3,$
$T_{0}:=(|z_{1}^{(0)}|+|z_{2}^{(0)}|)^{-1}$, $\nu$ is some large
defined in (~\ref{zeromodeestimate}) below, recall the definitions
of $z_{1}^{(0)}$ and $z_{2}^{(0)}$ in Main Theorem
~\ref{maintheorem1}.

Our next result is
\begin{theorem}\label{THM:formulascalar}
The functions $\dot{\lambda}$ and $\dot{\gamma}$ have the following
expansions in the parameters $y$ and $\bar{y}$:
\begin{equation}\label{expanlambda}
\dot{\lambda}=\sum_{2\leq m+n\leq
2N+1}\Lambda_{mn}(\lambda)y^{m}\bar{y}^{n}+Remainder,
\end{equation}
where $\Lambda_{mn}(\lambda)=\bar{\Lambda}_{nm}(\lambda),$ and
$\Lambda_{m,n}(\lambda)$ is purely imaginary for $m,n\leq N$,
(therefore $\Lambda_{mm}(\lambda)=0$ for $m\leq N$);
\begin{equation}\label{expangamma}
\dot{\gamma}=\sum_{2\leq m+n\leq
2N+1}\Gamma_{mn}(\lambda)y^{m}\bar{y}^{n}+Remainder,
\end{equation}
where $\Gamma_{mn}(\lambda)=\bar\Gamma_{nm}(\lambda),$ and
$\Gamma_{mn}(\lambda)$ is real for $m,n\leq N$. The terms
$Remainder$ are not the same in the equations above, but both admit
the estimate (~\ref{remainder}).
\end{theorem}
\subsection{Proof of Theorems ~\ref{expansion} and
~\ref{THM:formulascalar}}\label{proofexpansion} In this subsection
we prove Theorems ~\ref{expansion} and ~\ref{THM:formulascalar}. We
divide the proof into three steps. The following lemma will be used
repeatedly to prove the admissibility of the function
$R_{m,n}(\lambda)$.
\begin{lemma}\label{LM:admissibility}
If $K_{1}$ is a vector-function from $\mathbb{R}^{d}$ to $
\mathbb{C}^{2}$ such that $iK_{1}$ is admissible, then the vector
function
$$K_{2}:=(L(\lambda)-i\mu)^{-1}P_{c}K_{1}$$ is admissible for any $\mu\in (-\lambda,\lambda).$
\end{lemma}
\begin{proof}
First by Equation (~\ref{eq:concreteform}) we observe that
$iP_{c}K_{1}$ is admissible. Then the computation
$$
\begin{array}{lll}
\bar{K}_{2}&=&(L(\lambda)+i\mu)^{-1}P_{c}\bar{K}_{1}\\
&=&-(L(\lambda)+i\mu)^{-1}\sigma_{3}P_{c}K_{1}\\
&=&\sigma_{3}(L(\lambda)-i\mu)^{-1}P_{c}K_{1}\\
&=&\sigma_{3}K_{2},
\end{array}
$$ where $\sigma_{3}:=\left(
\begin{array}{lll}
1&0\\
0&-1
\end{array}
\right),$ implies that $K_{2}$ is admissible.
\end{proof}
\subsubsection{The first step: $z$-expansion}
In this sub-subsection we prove the following proposition.
\begin{proposition}\label{firststep}
For any $k=2,\cdot\cdot\cdot,N$
\begin{equation}\label{eq:decom}
\vec{R}=\sum_{2\leq m+n\leq
k}\tilde{R}_{m,n}(\lambda)z^{m}\bar{z}^{n}+\tilde{R}_{k}
\end{equation}
where the functions $\tilde{R}_{m,n}$ are admissible, and the
remainder $\tilde{R}_{k}$ satisfies the equation
\begin{equation}\label{eq:equationr6}
\begin{array}{lll}
\frac{d}{dt}\tilde{R}_{k}&=&L(\lambda)\tilde{R}_{k}+\dot\gamma
P_{c}J\tilde{R}_{k}+\dot{\lambda}P_{c\lambda}\tilde{R}_{k}+\tilde{A}_{k}(z,\bar{z})\tilde{R}_{k}\\
& &+\displaystyle\sum_{k+1\leq m+n\leq 2N
}R^{(1)}_{m,n}z^{m}\bar{z}^{n}+N_{k}(\tilde{R}_{k},z)+Remainder_{1}
\end{array}
\end{equation} where
the term $N_{k}(\tilde{R}_{k},z)$ contains all the nonlinear terms
in $\tilde{R}_{k}$ and, for $k=N$, is bounded as
$$
 \|N_{N}(\tilde{R}_{N},z)\|_{1}+\|N_{N}(\tilde{R}_{N},z)\|_{\mathcal{H}^{l}}\leq c
 (T_{0}+t)^{-\frac{2N+3}{2N}}[Y^{2}\mathcal{R}_{1}^{6}+\mathcal{R}_{1}^{5}\mathcal{R}_{2}^{2}],
$$
$$
 \begin{array}{lll}
 &
 &\|(-\Delta+1)^{\frac{l}{2}}N_{N}(\tilde{R}_{N},z)\|_{1}+\|N_{N}(\tilde{R}_{N},z)\|_{\mathcal{H}^{l}}\\
 &\leq&c(T_{0}+t)^{-\frac{2N+2}{2N}}(Y^{2}\mathcal{R}_{1}^{2}\mathcal{R}_{2}^{2}+Y^{2}\mathcal{R}_{1}^{3}\mathcal{R}_{2}^{3}+\mathcal{R}_{1}^{3}\mathcal{R}_{2}^{4})
 \end{array}
$$ with $l:=[\frac{d}{2}]+3,$ the functions $iR^{(1)}_{m,n}$ are admissible, smooth, and decay exponentially
fast at $\infty$, and the $(2\times 2)$-matrix function
$\tilde{A}_{k}(z,\bar{z})$ is bounded in the matrix norm as
$$|\tilde{A}_{k}(z,\bar{z})|\leq
c|z||e^{-\epsilon_{0}|x|},$$ and the function $Remainder_{1}$
satisfies the estimate
\begin{equation}\label{remainder1}
|Remainder_{1}|\leq c|z|^{2N+1}e^{-\epsilon_{0}|x|}.
\end{equation}
\end{proposition}
\begin{proof}
We prove the theorem by induction in $k$. Thus we first consider the
case $k=2.$ If we let
$$\tilde{R}_{2,0}(\lambda):=\frac{1}{4}[L(\lambda)-2i\epsilon(\lambda)]^{-1}P_{c}f^{'}((\phi^{\lambda})^{2})\phi^{\lambda}\left(
\begin{array}{lll}
2i\eta\xi\\
3\xi^{2}+\eta^{2}
\end{array}
\right),$$
$$\tilde{R}_{0,2}(\lambda):=\bar{\tilde{R}}_{2,0}(\lambda),$$
$$\tilde{R}_{11}(\lambda):=\frac{1}{2}L(\lambda)^{-1}P_{c}f^{'}((\phi^{\lambda})^{2})\phi^{\lambda}\left(
\begin{array}{lll}
0\\
3\xi^{2}-\eta^{2}
\end{array}
\right)$$ and
$$\tilde{R}_{2}:=\vec{R}-\displaystyle\sum_{m+n=2}z^{m}\bar{z}^{n}\tilde{R}_{m,n}(\lambda),$$
then the functions $\tilde{R}_{m,n}(\lambda),$ $m+n=2,$ are
admissible by Lemma ~\ref{LM:admissibility} and $\tilde{R}_{2}$
satisfies the equation (~\ref{eq:equationr6}) when $k=2.$ Thus we
obtain the first step of induction.

Now assume (~\ref{eq:decom}) holds for some $2\leq k-1< N$ and prove
it for $k.$ Define the term $\tilde{R}_{k}$ by (~\ref{eq:decom}). We
claim that $\tilde{R}_{k}$ satisfies the following equation:
\begin{equation}\label{equaLR}
\begin{array}{lll}
\frac{d}{dt}\tilde{R}_{k}&=&[L(\lambda)+\dot\gamma
P_{c}J+\dot{\lambda}P_{c\lambda}
+\tilde{A}_{k}(z,\bar{z})]\tilde{R}_{k}\\
& &+\displaystyle\sum_{2\leq m+n\leq
2N}z^{m}\bar{z}^{n}F_{m,n}+P_{c}JN_{k}(\tilde{R}_{k},z)+Remainder_{1}
\end{array}
\end{equation}
where for $m+n\leq k$
$$F_{m,n}:=[L(\lambda)-i\epsilon(\lambda)(m-n)]\tilde{R}_{mn}(\lambda)+P_{c}f_{m,n}(\lambda),$$
with functions $f_{m,n}(\lambda)$ having the following properties
\begin{enumerate}
\item[(A)] the functions $f_{m,n}(\lambda)$ depend on $\tilde{R}_{m',n'}(\lambda)$
 with $m'+n'<m+n$;
\item[(B)] $if_{m,n}(\lambda)$ are admissible, smooth and decays
exponentially fas,t provided that $\tilde{R}_{m',n'}(\lambda)$ are
admissible, smooth and decay exponentially fast for all pairs
$(m',n')$ satisfying $m'+n'<m+n$.
\end{enumerate}
We prove this claim below. Recall that if $|m-n|\leq N,$ then
$i\epsilon(\lambda)(m-n)\not\in \sigma(L(\lambda))$ and therefore
the operators
\begin{equation}\label{Invers}
L(\lambda)-i\epsilon(\lambda)(m-n):P_{c}\mathcal{L}^{2}\rightarrow
P_{c}\mathcal{L}^{2}
\end{equation} are invertible. Hence by Lemma
~\ref{LM:admissibility} the equations $F_{m,n}(\lambda)=0,$ $m+n\leq
k\leq N,$ have unique solutions with the property that
$\tilde{R}_{m,n}(\lambda)$ are admissible if $if_{m,n}(\lambda)$ are
admissible. By Claim (B), $if_{m,n}(\lambda)$ are admissible if
$\tilde{R}_{m',n'}(\lambda)$, $m'+n'<m+n,$ are admissible. This and
the induction in $k$ show the admissibility of
$\tilde{R}_{m,n}(\lambda)$ for $m+n\leq N.$

What is left is to prove the claims above. To this latter end we
plug decomposition (~\ref{eq:decom}) into Equation (~\ref{eq:R1R22})
to obtain $$
\begin{array}{lll}
\frac{d}{dt}\tilde{R}_{k}&=&L(\lambda)\tilde{R}_{k}+F(\vec{R},z)+\displaystyle\sum_{2\leq
m+n\leq k}z^{m}\bar{z}^{n}[L(\lambda)-i(m-n)\epsilon(\lambda)]\tilde{R}_{m,n}(\lambda)\\
& &-\displaystyle\sum_{2\leq m+n\leq k}\tilde{R}_{m,n}(\lambda)[\frac{d}{dt}z^{m}\bar{z}^{n}-i(m-n)\epsilon(\lambda)z^{m}\bar{z}^{n}]\\
& &-\dot\lambda\displaystyle\sum_{2\leq m+n\leq
k}\partial_{\lambda}\tilde{R}_{m,n}(\lambda)z^{m}\bar{z}^{n}
\end{array}
$$ where the term $F(\vec{R},z)$ is given by $$
\begin{array}{lll}
F(\vec{R},z)&:=&\dot\lambda P_{c\lambda}\vec{R}+\dot\gamma P_{c}
J\vec{R}+P_{c}J\vec{N}(\vec{R},z)\\&+&\frac{1}{2}\dot\gamma
P_{c}[z\left(
\begin{array}{lll}
-i\eta\\
\xi
\end{array}
\right)+\bar{z}\left(
\begin{array}{lll}
i\eta\\
\xi
\end{array}
\right)]-\frac{1}{2}\dot\lambda P_{c}[z\left(
\begin{array}{lll}
\xi_{\lambda}\\
-i\eta_{\lambda}
\end{array}
\right)+\bar{z}\left(
\begin{array}{lll}
\xi_{\lambda}\\
i\eta_{\lambda}
\end{array}
\right)].
\end{array}
$$
Moreover $J\vec{N}(\vec{R},z):= J\left(
\begin{array}{lll}
Re N(R,z_{1},z_{2})\\
Im N(R,z_{1},z_{2})
\end{array}
\right)$ admits the expansion
\begin{equation}\label{NonExpan}
J\vec{N}(\vec{R},z)=\sum_{2\leq m+n\leq
2N}z^{m}\bar{z}^{n}N_{mn}(\lambda)+\tilde{A}_{k}(z,\bar{z})\tilde{R}_{k}+N_{k}(\tilde{R}_{k},z)+Remainder_{1}
\end{equation}
where $N_{m,n}(\lambda):\mathbb{R}^{3}\rightarrow \mathbb{C}^{2}$,
$\tilde{A}_{k}(z,\bar{z})$ as above, $N_{k}(\tilde{R}_{k},z)$
contains all the nonlinear terms in $\tilde{R}_{k}$ and the term
$Remainder_{1}$ has the same estimate as in Equation
(~\ref{remainder1}).

By Equations (~\ref{nonlinear}), (~\ref{eq:z2}) and
(~\ref{eq:lambda}) for $\vec{N}(\vec{R},z)$, $\dot{z},$
$\dot\lambda$ and $\dot\gamma$, to prove the claim it suffices to
prove that given $(m,n),$ the function $iN_{m,n}(\lambda)$ in
Equation (~\ref{NonExpan}) is admissible if
$\tilde{R}_{m',n'}(\lambda)$ are admissible for all $m'+n'<m+n$, and
depends only on $\tilde{R}_{m',n'}(\lambda),$ $m'+n'<m+n.$ The proof
of this sufficient condition is tedious and not hard, thus omitted.
\begin{enumerate}
\item[(1)]
To prove the admissibility of $iN_{m,n}(\lambda)$ we use the
definition of $N(R,z_{1},z_{2})$ in Equation (~\ref{nonlinear})
again. Note that if $f_{m,n}(\lambda)$ and $F_{m',n'}(\lambda)$ are
real and admissible functions, respectively, then the
vector-function $f_{m,n}F_{m',n'}(\lambda)$ is admissible. Therefore
it is sufficient to prove that if $\tilde{R}_{m',n'}(\lambda),$
$m'+n'<m+n,$ are admissible, then we have the expansion
\begin{equation}\label{nonlinearExpan}
\begin{array}{lll}
f(|\phi^{\lambda}+z_{1}\xi+iz_{2}\eta+R|^{2})&=&f((\phi^{\lambda})^{2})+\displaystyle\sum_{1\leq
m+n\leq
2N}f_{m,n}(\lambda)z^{m}\bar{z}^{n}+g(\tilde{R}_{k})\\
& &+Remainder_{1}
\end{array}
\end{equation}
where the functions $f_{m,n}(\lambda)$ are real, $g$ collects all
the linear and nonlinear terms containing $\tilde{R}_{k}$; it obeys
the estimate
$$\|g(\tilde{R}_{k})\|_{2}\leq
c(|z|^{4}\|e^{-\epsilon_{0}|x|}\tilde{R}_{k}\|_{2}+\|\tilde{R}^{3}_{k}\|_{2})$$
for some constant $\epsilon_{0}>0,$ and $Remainder_{1}$ satisfies
the estimate (~\ref{remainder1}). Indeed, let $\vec\phi:=\left(
\begin{array}{lll}
\phi^{\lambda}+\frac{z+\bar{z}}{2}\xi\\
\frac{z-\bar{z}}{2i}\eta
\end{array}
\right),$ then
$$
\begin{array}{lll}
|\phi^{\lambda}+z_{1}\xi+iz_{2}\eta+R|^{2}&=&|\vec\phi+\vec{R}|^{2},
\end{array}
$$ where,
recall that $\vec{R}:=\left(
\begin{array}{lll}
ReR\\
ImR
\end{array}
\right).$ Let $\vec{R}_{k}:=\displaystyle\sum_{2\leq m+n\leq
k}\tilde{R}_{m,n}(\lambda)z^{m}\bar{z}^{n}$ be a real function in
Equation (~\ref{eq:decom}). We find
$$|\vec\phi+\vec{R}|^{2}=\vec\phi\cdot\vec\phi+2\vec\phi\cdot\vec{R}_{k}+\vec{R}_{k}\cdot\vec{R}_{k}+2(\vec\phi+\vec{R}_{k})\cdot\tilde{R}_{k}+\tilde{R}_{k}\cdot\tilde{R}_{k}.$$
Since the vector-functions $\vec{\phi}$ and $\vec{R}_{k}$ have
finite $z-$expansions with admissible coefficients, the first three
functions on the right hand side have finite $z-$expansions with
real coefficients. Moreover the expansion for $\vec\phi
\cdot\vec\phi$ starting with the term $(\phi^{\lambda})^{2}.$
Expanding the function
$f(|\phi^{\lambda}+z_{1}\xi+iz_{2}\eta+R|^{2})$ around
$(\phi^{\lambda})^{2}$ to the $2N-$th order, we have Equation
(~\ref{nonlinearExpan}).
\item[(2)]
The fact that $N_{mn}(\lambda)$ depends only on the terms
$\tilde{R}_{m',n'},$ $m'+n'<m+n,$ follows from by the computation in
Statement (1) above.
\end{enumerate}
\end{proof}
We plug the expansion of the function $\vec{R}$ into (~\ref{eq:z2})
and (~\ref{eq:lambda}) to obtain the following expansions for
$\dot\lambda,\ \dot\gamma$ and $\dot{z}:$
\begin{corollary}
\begin{equation}\label{eq:firstdecom}
\begin{array}{lll}
\dot\gamma&=&\displaystyle\sum_{2\leq
m+n\leq 2N+1}\tilde\Gamma_{m,n}(\lambda)z^{m}\bar{z}^{n}+\tilde{l}_{\gamma}(\tilde{R}_{N})+Remainder;\\
\dot\lambda &=&\displaystyle\sum_{2\leq m+n\leq 2N+1}\tilde\Lambda_{m,n}(\lambda)z^{m}\bar{z}^{n}+\tilde{l}_{\lambda}(\tilde{R}_{N})+Remainder;\\
\end{array}
\end{equation}
\begin{equation}\label{eq:equationz}
\dot{z}=i\epsilon(\lambda)z+\displaystyle\sum_{2\leq m+n\leq
2N+1}\tilde{Z}_{m,n}(\lambda)z^{m}\bar{z}^{n}+\tilde{l}_{z}(\tilde{R}_{N})+Remainder
\end{equation}
where, $\tilde{l}_{\gamma},\ \tilde{l}_{\lambda}$ and
$\tilde{l}_{z}$ are linear functionals having the estimates
\begin{equation}\label{eq:linestimate}
|\tilde{l}_{\gamma}(g)|,\ |\tilde{l}_{\lambda}(g)|,\
|\tilde{l}_{z}(g)|\leq c|z|\|e^{-\epsilon_{0}|x|}g\|_{2},
\end{equation} the term $Remainder$ admits the same estimate as in
(~\ref{remainder}), the coefficients $\tilde\Gamma_{m,n}(\lambda)$
are real, and $\tilde\Lambda_{m,n}(\lambda)$ and
$\tilde{Z}_{m,n}(\lambda)$ are purely imaginary.
\end{corollary}
The proof is straightforward by Proposition ~\ref{firststep},
Equations (~\ref{eq:lambda}) (~\ref{eq:z2}) and the properties of
the term $J\vec{N}(\vec{R},z)$ in (~\ref{NonExpan}), and thus is
omitted.
\subsubsection{The second step: changing variables}
In the second step we transform $z$ to a parameter $y$ which
satisfies a simpler different equation.
\begin{proposition}\label{PRO:transform}
There exists a polynomial $P(z,\bar{z})$ with real coefficients and
the smallest degree $\geq 2$, such that if we define
$y:=z+P(z,\bar{z})$ then
\begin{equation}\label{eq:tranform}
\begin{array}{lll}
\dot{y}=i\epsilon(\lambda)y+\displaystyle\sum_{2\leq m+n\leq
2N+1}Y_{m,n}(\lambda)y^{m}\bar{y}^{n}+l_{y}(\tilde{R}_{N})
+Remainder
\end{array}
\end{equation} where the coefficients $Y_{m,n}(\lambda)$
are purely imaginary, especially $Y_{m,n}=0$ if $m+n\leq N$ and
$m\not=n+1,$ $l_{y}$ is a linear functional satisfying the estimate
$$|l_{y}(g)|\leq c|y|\|e^{-\epsilon_{0}|x|}g\|_{2},$$ and the term
$Remainder$ admits the estimate (~\ref{remainder}).
\end{proposition}
\begin{proof}
We show how to construct the polynomial $P(z,\bar{z})$. We rewrite
Equation (~\ref{eq:equationz}) as
\begin{equation}\label{firstPeel}
\begin{array}{lll}
&
&\partial_{t}(z-\displaystyle\sum_{m+n=2}\frac{\tilde{Z}_{m,n}(\lambda)}{i(m-n-1)\epsilon(\lambda)}z^{m}\bar{z}^{n})\\
&=&i\epsilon(\lambda)[z-\displaystyle\sum_{m+n=2}\frac{\tilde{Z}_{m,n}(\lambda)}{i(m-n-1)\epsilon(\lambda)}z^{m}\bar{z}^{n}]+D+\tilde{l}_{z}(\tilde{R}_{N})+Remainder,
\end{array}
\end{equation} where the linear functional $\tilde{l}_{z}$ satisfies the same estimate as in (~\ref{eq:linestimate}), the term $D$ is given by
$$
\begin{array}{lll}
D&:=&-\frac{d}{dt}\displaystyle\sum_{m+n=2}\frac{\tilde{Z}_{m,n}(\lambda)}{i(m-n-1)\epsilon(\lambda)}z^{m}\bar{z}^{n}+\displaystyle\sum_{m+n=2}\frac{\tilde{Z}_{m,n}(\lambda)}{m-n-1}z^{m}\bar{z}^{n}\\
& &+\displaystyle\sum_{2\leq m+n\leq
2N+1}\tilde{Z}_{m,n}(\lambda)z^{m}\bar{z}^{n}.
\end{array}
$$ Take the time derivative on the right hand side and use Equations (~\ref{eq:firstdecom}) and (~\ref{eq:equationz}) to get $$D=\displaystyle\sum_{3\leq m+n\leq 2N+1}a^{(1)}_{m,n}(\lambda)z^{m}\bar{z}^{n}+Remainder.$$
Since $\tilde{Z}_{m,n}(\lambda)$ and $\tilde{\Lambda}_{m,n}$ are
purely imaginary we see that $a^{(1)}_{m,n}(\lambda)$ are purely
imaginary. Now define
\begin{equation}\label{P1}
P_{1}(z,\bar{z}):=-\sum_{m+n=2}\frac{\tilde{Z}_{mn}(\lambda)}{i(m-n-1)\epsilon(\lambda)}z^{m}\bar{z}^{n}
\end{equation}
and $y_{1}:=z+P_{1}(z,\bar{z}).$ We observe that
$\frac{\tilde{Z}_{m,n}(\lambda)}{i(m-n)\epsilon(\lambda)}$ are real.
Then Equation (~\ref{firstPeel}) yields
$$\dot{y}_{1}=i\epsilon(\lambda)y_{1}+\sum_{3\leq m+n\leq
2N+1}a_{m,n}^{(2)}(\lambda)y_{1}^{m}\bar{y}_{1}^{n}+l_{y_{1}}(\tilde{R}_{N})+Remainder$$
where $a_{m,n}^{(2)}(\lambda)$ are purely imaginary, and the term
$Remainder$ has the same estimate as in (~\ref{remainder}).

Next we remove the terms with $m+n=3$ and $m\not= n+1$ and so forth
arriving at the end at Equation (~\ref{eq:tranform}).
\end{proof}
We invert the relations $y=z+P(z,\bar{z})$ and
$\bar{y}=\bar{z}+\bar{P}(z,\bar{z})$ and express the variables $z$
and $\bar{z}$ as power series in $y$ and $\bar{y}.$ Plug the result
into (~\ref{eq:firstdecom}) for $\dot\gamma$ and $\dot\lambda$ and
into Equations (~\ref{eq:decom}) for $\vec{R}$ to obtain the
expansions
$$
\dot\gamma=\displaystyle\sum_{2\leq m+n\leq
2N+1}\Gamma^{(1)}_{m,n}(\lambda)y^{m}\bar{y}^{n}+l_{\gamma}(R_{N})+Remainder,
$$ $$
\dot\lambda =\displaystyle\sum_{2\leq m+n\leq
2N+1}\Lambda^{(1)}_{m,n}(\lambda)y^{m}\bar{y}^{n}+l_{\lambda}(R_{N})+Remainder;
$$
$$\vec{R}=\sum_{2\leq m+n\leq k}R_{m,n}(\lambda)y^{m}\bar{y}^{n}+R_{k},$$
where $\Gamma^{(1)}_{m,n}(\lambda)$ are real,
$\Lambda^{(1)}_{m,n}(\lambda)$ are purely imaginary, $2\leq k\leq
N$, the linear functionals $l_{\gamma},\ l_{\lambda}$ satisfy the
estimate
$$|l_{\gamma}(g)|, |l_{\lambda}(g)|\leq
c|y|\|e^{-\epsilon_{0}|x|}g\|_{2},$$ $R_{m,n}$ are admissible, and
$R_{k}$ satisfies the equation
\begin{equation}\label{transformN}
\begin{array}{lll}
\frac{d}{dt}R_{k}&=&L(\lambda)R_{k}+\dot\gamma P_{c}J
R_{k}+\dot\lambda
P_{c\lambda}R_{k}+A_{k}(y,\bar{y})R_{k}\\
& &+\displaystyle\sum_{N+1\leq m+n\leq
2N+1}iR^{(k)}_{m,n}(\lambda)y^{m}\bar{y}^{n}+N_{k}(R_{k},y)+Remainder_{1},
\end{array}
\end{equation}
with the functions $R^{(k)}_{m,n}(\lambda)$ admissible, and
$N_{N}(R_{N},y)$ satisfying the estimates
(~\ref{remainderestimate})-(~\ref{remainderestiamte2}) and the
operator $A_{k}(y,\bar{y})$ have the same estimates as that in
(~\ref{eq:equationr6}), and the terms $Remainder$ and
$Remainder_{1}$ admit the same estimates as in (~\ref{remainder})
and (~\ref{remainder1}), respectively. Note that the polynomial
$P_{1}$ in (~\ref{P1}) has real coefficients and therefore the
expansion of $z$ and $\bar{z}$ in powers of $y$ and $\bar{y}$ has
real coefficients also. Since a product of real and admissible
functions is admissible we conclude that the coefficients
$R_{m,n}(\lambda)$ are also admissible.

The above relations prove Theorem ~\ref{expansion}, except for
(~\ref{expanz2}), for $2\leq k\leq N.$ The proof for $N<k\leq 2N$ is
more difficult since $i\epsilon(\lambda)(m-n)$ in (~\ref{Invers})
might be in the spectrum of $L(\lambda).$ This is done in the next
step.
\subsubsection{The Third Step: $N<k\leq 2N$. Completion of the Proof of Theorems ~\ref{expansion} and ~\ref{THM:formulascalar}}
This is the last and more involved step. As in the first step we
determine the coefficients $R_{m,n}(\lambda)$ by solving the
equations
$$[L(\lambda)-i\epsilon(\lambda)(m-n)]R_{m,n}(\lambda)=-P_{c}f_{m,n}(\lambda)$$ for certain functions $f_{m,n}(\lambda)$ (see below). Recall that the number $N$ is defined by the properties
$$i\epsilon(\lambda)(m-n)
\begin{array}{lll}
\not\in\sigma(L(\lambda))\ \text{if}\ |m-n|\leq N,\\
\in\sigma(L(\lambda))\ \text{if}\ |m-n|> N.
\end{array}
$$
Thus we sort out the pairs $(m,n)$ into "non-resonant pairs"
satisfying $|m-n|\leq N$ and "resonant pairs" satisfying $|m-n|> N.$
For "non-resonant" pairs the operators
$$L(\lambda)-i\epsilon(\lambda)(m-n): P_{c}\mathcal{L}^{2}\rightarrow
P_{c}\mathcal{L}^{2}$$ are invertible and for resonant pairs they
are not (one has to change spaces in the latter case). In the first
two steps we expanded in $z$ and $\bar{z}$ (and in $y$ and
$\bar{y}$) until $m+n\leq N$ and consequently all the pairs,
$(m,n)$, involved were non-resonant ones. Now, for $k>N,$ our
expansion involves pairs $(m,n)$ with $m+n>N,$ which include
resonant pairs. What we want to show now is that for the subsets of
pairs $(m,n)$, $m+n>N,$ determined by the inequality
$$m,n\leq N,$$ our analysis will involve only "non-resonant" pairs and we will
be able to prove the admissibility of the coefficients
$R_{m,n}(\lambda)$ in this case.

\begin{definition}
Suppose that $(m_{1},n_{1})$ and $(m_{2},n_{2})$ are two pairs of
nonnegative integers. Then $(m_{1},n_{1})<(m_{2},n_{2})$ if
$m_{1}\leq m_{2},$ $n_{1}\leq n_{2}$ and $(m_{1},n_{1})\not=
(m_{2},n_{2})$; and $(m_{1},n_{1})\leq(m_{2},n_{2})$ if $m_{1}\leq
m_{2},$ $n_{1}\leq n_{2}.$
\end{definition}
To prove Theorem ~\ref{expansion} for $N+1\leq k\leq 2N$ we proceed
as in the proof of the first step.
\begin{lemma}\label{expandk}
Let $N<k\leq 2N$. Then the remainder term $R_{k}$ in Equation
(~\ref{expansionR1}) satisfies the equation
$$
\begin{array}{lll}
\frac{d}{dt}R_{k}&=&L(\lambda)R_{k}+P_{k}(y,\bar{y})R_{k}+\displaystyle\sum_{2\leq
m+n\leq
k}y^{m}\bar{y}^{n}F_{m,n}(\lambda)\\
& &+P_{c}J\vec{N}_{N}(\tilde{R}_{N},y,\bar{y})+F_{k}(y,\bar{y}),
\end{array}
$$ where $P_{k}(y,\bar{y}),$
$N_{N}(R_{N},y,\bar{y})$ and $F_{k}(y,\bar{y})$ are described in
Theorem ~\ref{expansion}; $F_{mn}(\lambda)$ are the functions
defined as
$$F_{m,n}:=[L(\lambda)-i\epsilon(\lambda)(m-n)]R_{m,n}(\lambda)+P_{c}f_{m,n}(\lambda)$$
where the functions $f_{m,n}(\lambda)$ have the following
properties:
\begin{enumerate}
\item[(A)] if $m,n\leq N$ and all the terms $R_{m_{1},n_{1}}(\lambda),$ $(m_{1},n_{1})<(m,n),$ are
admissible then $if_{m,n}$ is admissible;
\item[(B)] if $\max\{m,n\}>N$ then $f_{m,n}(\lambda)$ is of the form
(~\ref{eq:unusual}).
\end{enumerate}
Moreover we have the following expansions for $\dot{y},$
$\dot{\lambda}$ and $\dot{\gamma}:$
\begin{equation}\label{yk}
\dot{y}=i\epsilon(\lambda)y+\sum_{2\leq m+n\leq 2N+1
}\Theta_{m,n}(\lambda)y^{m}\bar{y}^{n}+l_{y}^{(k)}(R_{k})+Remainder,
\end{equation}
\begin{equation}\label{lambdak}
\dot\lambda=\sum_{2\leq m+n\leq
2N+1}\Lambda_{m,n}(\lambda)y^{m}\bar{y}^{n}+l_{\lambda}^{(k)}(R_{k})+Remainder
\end{equation} and
\begin{equation}\label{gammak}
\dot\gamma=\sum_{2\leq m+n\leq
2N+1}\Gamma_{m,n}(\lambda)y^{m}\bar{y}^{n}+l_{\gamma}^{(k)}(R_{k})+Remainder
\end{equation}
where $\Theta_{m,n}(\lambda)=0$ for $m+n\leq N$ and $m\not= n+1$,
$l_{\lambda}^{(k)},$ $l_{y}^{(k)}$ and $l_{\gamma}^{(k)}$ are linear
functionals of the first-order in $y$ satisfying the estimates
\begin{equation}\label{eq:linEstK}
|l_{\lambda}^{(k)}(g)|,\ |l_{y}^{(k)}(g)|, \
|l_{\gamma}^{(k)}(g)|\leq c|y|\|e^{-\epsilon_{0}|x|}g\|_{2},
\end{equation} $Remainder$ obeys the
estimate in Equation (~\ref{remainder}). Moreover, if the functions
$R_{m_{1},n_{1}}(\lambda)$ are admissible for all pairs
$(m_{1},n_{1})<(m,n)$ with $m,n\leq N$ and $m+n\leq k$, then
$\Lambda_{m,n}(\lambda)$ and $\Theta_{m,n}(\lambda)$ are purely
imaginary and $\Gamma_{m,n}(\lambda)$ are real.
\end{lemma}
We prove this lemma in Appendix ~\ref{AP:Lemma}. Meantime we proceed
with the proof of Theorem ~\ref{expansion}. We determine the
coefficients $R_{m,n}(\lambda),$ $m+n\leq k,$ by solving the
equations $F_{mn}(\lambda)=0$, i.e.
\begin{equation}\label{findSolution}
[L(\lambda)-i\epsilon(\lambda)(m-n)]R_{mn}(\lambda)=-P_{c}f_{mn}(\lambda).
\end{equation}
By Lemma ~\ref{LM:admissibility} we have that $R_{m,n}(\lambda)$
solving Equation (~\ref{findSolution}) is admissible for $m,n\leq
N$, (and hence $|m-n|\epsilon(\lambda)< \lambda$), if so is
$if_{m,n}(\lambda).$ By Property (A) in Lemma ~\ref{expandk},
$if_{m,n}(\lambda)$ is admissible if so are $R_{m',n'}(\lambda)$
with $(m',n')<(m,n).$ Thus if $R_{m',n'}(\lambda)$ is admissible for
every $(m',n')<(m,n),$ then so is $R_{m,n}(\lambda)$. Since
$R_{m,n}(\lambda),\ m+n\leq N$ are admissible, we have by induction
in $(m',n')$ that $R_{m,n}(\lambda), \ m,n\leq N,$ are admissible.
This proves (~\ref{expansionR1}) with (RA) and (RB). Property (RC)
follows from Lemma ~\ref{expandk} and the equations
$F_{m,n}(\lambda)=0,\ 2\leq m+n\leq k$.

Furthermore, when $k=2N$, we have by (~\ref{eq:linEstK}) above that
$$|l_{\lambda}^{(2N)}(R_{2N})|,\ |l_{y}^{(2N)}(R_{2N})|, \
|l_{\gamma}^{(2N)}(R_{2N})|\leq
c|y|\|e^{-\epsilon_{0}|x|}R_{2N}\|_{2}.$$ Moreover, since
$$|y(t)|\|e^{-\epsilon_{0}|x|}R_{2N}(t)\|_{2}\leq
c(1+t)^{-\frac{2N+2}{2N}}Y(t)\mathcal{R}_{1}(t),$$ where the
estimating functions $Y$ and $\mathcal{R}_{1}$ are defined in
(~\ref{majorant}), the terms $l_{\lambda}^{(2N)}(R_{2N}),\
l_{y}^{(2N)}(R_{2N})$ and $l_{\gamma}^{(2N)}(R_{2N})$ obey the
estimates in (~\ref{eq:linEstK}) and therefore can be placed into
$Remainder$. Hence the equations for $\dot{y}$, $\dot{\lambda}$ and
$\dot{\gamma}$ in Lemma ~\ref{expandk} imply the corresponding
equations given in (~\ref{expanz2}) and Theorem
~\ref{THM:formulascalar}.
\begin{flushright}
$\square$
\end{flushright}
\section{Estimates on $\lambda$}\label{lambdacongenvent}
In this section we obtain an estimate which, together with estimates
on $Y(T)$ and $R_{j}(T), \ j=1,2,3,$ obtained in Section
~\ref{ProveMain}, will imply the convergence of the parameter
$\lambda(t)$ as $t\rightarrow \infty.$
\begin{proposition}\label{Pro:lamCon}
There exists a constant $c$ such that for any $t$ and $T$ such that
$t\leq T$
\begin{equation}\label{convergencelambda}
|\lambda(t)-\lambda(T)|\leq
c(T_{0}+t)^{-\frac{1}{2N}}(Y(T)+\mathcal{R}_{1}(T))^{2}.
\end{equation}
\end{proposition}
\begin{proof}
First we note that Equation (~\ref{expanlambda}) does not imply
directly Estimate (~\ref{convergencelambda}). To obtain
(~\ref{convergencelambda}) we transform $y$ as
\begin{proposition}\label{leadingorder}
There exists a transformation $y$ to $\beta$ s.t.
$\beta=y+O(|y|^{2})$ and
\begin{equation}\label{converlambdaz}
\frac{d}{dt}[\lambda-\displaystyle\sum_{2\leq m+n\leq
2N+1}a_{m,n}(\lambda)\beta^{m}\bar{\beta}^{n}]=Remainder,
\end{equation} where, $a_{m,n}(\lambda):\mathbb{R}^{+}\rightarrow
\mathbb{C}$ and the $Remainder$ satisfies Estimate
(~\ref{remainder}).
\end{proposition}
This proposition will be proved in Subsection
~\ref{sec:proofleadingorder}. By Proposition ~\ref{leadingorder} we
have
$$
\begin{array}{lll}
& &|\lambda(t)-\displaystyle\sum_{2\leq m+n\leq
2N+1}a_{m,n}(\lambda(t))\beta^{m}\bar\beta^{n}(t)\\
& &-\lambda(T)+\displaystyle\sum_{2\leq m+n\leq
2N+1}a_{m,n}(\lambda(T))\beta^{m}\bar\beta^{n}(T)|\\
&=&|\int_{t}^{T}Remainder(s) ds|.
\end{array}
$$ By the estimate of $Remainder$ in
(~\ref{remainder}) we have that for any $t\leq T$
$$|\int_{t}^{T}Remainder(s) ds|\leq c(T_{0}+t)^{-\frac{1}{N}}(Y(T)+\mathcal{R}_{1}(T))^{2}.$$
By the definition of $Y$ in Equation (~\ref{majorant}) and the fact
that $\beta=y+O(|y^{2}|)$ we have $|\beta(t)|\leq
c(1+t)^{-\frac{1}{2N}}Y(t)$ for some constant $c.$ Therefore
(~\ref{convergencelambda}) follows.
\end{proof}
\subsection{Proof of Proposition ~\ref{leadingorder}}\label{sec:proofleadingorder}
\textbf{Below $Remainder$ signifies a function satisfying
(~\ref{remainder}).} We begin with
\begin{lemma}\label{transformz}
There exists a polynomial $$P_{1}(y,\bar{y})=\sum_{N+1\leq m+n\leq
2N+1}u_{mn}(\lambda)y^{m}\bar{y}^{n}$$ where the coefficients
$u_{m,n}$ are real for $m,n\leq N$, such that if we let
$$\beta:=y+P_{1}(y,\bar{y})$$ then
\begin{enumerate}
\item[(A)]
$\dot{\beta}=i\epsilon(\lambda)\beta+\displaystyle\sum_{1\leq n\leq
N}Y_{n}(\lambda)\beta^{n+1}\bar{\beta}^{n}+Remainder,$ with the
coefficients $Y_{n}(\lambda)$ purely imaginary for $n<N$;

\item[(B)] $\dot{\lambda}=\displaystyle\sum_{2\leq m+n\leq 2N+1}\lambda_{m,n}(\lambda)\beta^{m}\bar{\beta}^{n}+Remainder,$ with
$\lambda_{m,n}(\lambda)$ purely imaginary for any $m,n\leq N$.
\end{enumerate}
\end{lemma}
The proof of this lemma is given in Subsection
~\ref{prooftransformationz}. Note that Statement (A) is the same as
Statement (B) of Main Theorem ~\ref{maintheorem2}. We prove
Proposition ~\ref{leadingorder} by using inductions on the number
$k=m+n$. Suppose that for $1\leq k<2N+1$
$$\frac{d}{dt}[\lambda-\sum_{2\leq m+n\leq k}a_{mn}(\lambda)\beta^{m}\bar{\beta}^{n}]=\sum_{k+1\leq m+n\leq 2N+1}b_{m,n}(\lambda)\beta^{m}\bar{\beta}^{n}+Remainder,$$
where $b_{m,n}(\lambda)$ are purely imaginary for $m,n\leq N$. This
latter properties together with the fact that $\lambda$ is real
imply that $b_{nn}(\lambda)=0.$ Since $a_{m,n}(\lambda)=0$ for
$m+n=1$, the first step of induction, $k=1,$ is automatically true.

To remove the leading order from the right hand side of the last
equation we rewrite it as
\begin{equation}\label{remindpeellambda}
\begin{array}{lll}
& &\frac{d}{dt}[\lambda-\displaystyle\sum_{m+n\leq
k}a_{mn}(\lambda)\beta^{m}\bar{\beta}^{n}-\sum_{m+n=k+1}\frac{b_{m,n}(\lambda)}{i(m-n)\epsilon(\lambda)}\beta^{m}\bar{\beta}^{n}]\\
&=&B_{k+1}+\displaystyle\sum_{k+2\leq m+n\leq
2N+1}b_{m,n}(\lambda)\beta^{m}\bar{\beta}^{n}+Remainder.
\end{array}
\end{equation}
where
\begin{equation}\label{secondremainder}
B_{k+1}:=\displaystyle\sum_{m+n=k+1}b_{m,n}(\lambda)\beta^{m}\bar{\beta}^{n}-\frac{d}{dt}\sum_{m+n=k+1}\frac{b_{m,n}(\lambda)}{i(m-n)\epsilon(\lambda)}\beta^{m}\bar{\beta}^{n}.
\end{equation} Then the right hand side of Equation (~\ref{remindpeellambda}) is of
order $|\beta|^{k+2}.$

By the $k-$ step assumption the second term on the right hand side
of (~\ref{remindpeellambda}) is of the form required by the
$(k+1)-$step of the induction. Now we show that the first term on
the right hand side, $B_{k+1}$, is also of the right form, i.e.
$$B_{k+1}=\displaystyle\sum_{k+1\leq m+n\leq
2N+1}c_{m,n}(\lambda)\beta^{m}\bar{\beta}^{n}+Remainder$$ where the
coefficients $c_{m,n}(\lambda)$ are purely imaginary for $m,n\leq
N.$ Indeed, we expand the term (~\ref{secondremainder}) as
\begin{equation}\label{eq:industep}
\begin{array}{lll}
B_{k+1}&=&-\displaystyle\sum_{m+n=k+1}\dot{\lambda}\frac{d}{d\lambda}(\frac{b_{m,n}(\lambda)}{i(m-n)\epsilon(\lambda)})\beta^{m}\bar{\beta}^{n}\\
&
&-\displaystyle\sum_{m+n=k+1}\frac{b_{m,n}(\lambda)}{i(m-n)\epsilon(\lambda)}(\frac{d}{dt}\beta^{m}\bar{\beta}^{n}-i(m-n)\epsilon(\lambda)\beta^{m}\bar{\beta}^{n})+Remainder\\
&=&-\displaystyle\sum_{2\leq m'+n'\leq 2N+1}\sum_{m+n=k+1}\lambda_{m'n'}(\lambda)\frac{d}{d\lambda}(\frac{b_{m,n}(\lambda)}{i(m-n)\epsilon(\lambda)})\beta^{m+m'}\bar\beta^{n+n'}\\
& &-\displaystyle\sum_{1\leq n'\leq
N}\sum_{m+n=k+1}[m\frac{b_{m,n}Y_{n'}(\lambda)}{i(m-n)\epsilon(\lambda)}+n\frac{b_{m,n}\bar{Y}_{n'}(\lambda)}{i(m-n)\epsilon(\lambda)}]\beta^{m+n'}\bar\beta^{n+n'}\\
& &+Remainder.
\end{array}
\end{equation}
By the properties of $b_{m,n},\lambda_{m,n}$ and $Y_{n}$, we have
that if $m+m',\ n+n'\leq N$ then
$\lambda_{m'n'}(\lambda)\frac{d}{d\lambda}(\frac{b_{m,n}(\lambda)}{i(m-n)\epsilon(\lambda)})$
is purely imaginary; if $m+n',n+n'\leq N$ then
$m\frac{b_{m,n}Y_{n'}(\lambda)}{i(m-n)\epsilon(\lambda)}$ and
$n\frac{b_{m,n}\bar{Y}_{n'}(\lambda)}{i(m-n)\epsilon(\lambda)}$ are
purely imaginary.

Thus we proved that
$$
\begin{array}{lll}
& &\frac{d}{dt}[\lambda-\displaystyle\sum_{2\leq m+n\leq
k}a_{mn}(\lambda)\beta^{m}\bar{\beta}^{n}-\sum_{m+n=k+1}\frac{b_{m,n}(\lambda)}{i(m-n)\epsilon(\lambda)}\beta^{m}\bar{\beta}^{n}]\\
&=&\displaystyle\sum_{k+2\leq m+n\leq
2N+1}b^{(1)}_{m,n}(\lambda)\beta^{m}\bar\beta^{n}+Remainder
\end{array}
$$ where the coefficients $b^{(1)}_{m,n}(\lambda)$ are purely imaginary for $m,n\leq N.$
Thus the induction is complete. Taking $k=2N+1$ yields Equation
(~\ref{converlambdaz}).
\begin{flushright}
$\square$
\end{flushright}
\subsection{Proof of Lemma
~\ref{transformz}}\label{prooftransformationz} \textbf{Below
$Remainder$ signifies a term satisfying (~\ref{remainder}).} We
prove Statement (A) by induction. We define a set
\begin{equation}\label{eq:sets}
\mathcal{A}_{k}:=\{(m,n)|m,n\in \mathbb{Z}^{+},\ m+n=k, \
m\not=n+1\}.
\end{equation}
Suppose that for $N<k\leq 2N+1$ we found a transformation
$\beta_{k}=y+P_{1}^{(k)}(y,\bar{y})$ such that $\beta_{k}$ satisfies
the equation
$$
\begin{array}{lll}
\dot{\beta_{k}}&=&i\epsilon(\lambda)\beta_{k}+\displaystyle\sum_{n=1}^{N}\Theta_{n}(\lambda)\beta_{k}^{n+1}\bar{\beta}_{k}^{n}+\displaystyle\sum_{k\leq l\leq 2N+1}\displaystyle\sum_{(m,n)\in\mathcal{A}_{l}}\Theta_{m,n}(\lambda)\beta_{k}^{m}\bar{\beta}_{k}^{n}\\
& &+Remainder
\end{array}
$$
where $\Theta_{n}(\lambda)\equiv \Theta_{n,n}(\lambda)$ are purely
imaginary if $n< N$ and $\Theta_{m,n}(\lambda)$ are purely imaginary
for $m,n\leq N$. Note that by (~\ref{expanz2}) when $k=N+1$ the
equation above holds for $\beta_{k}=y$. Thus we have the first step
of the induction.

We have that
\begin{equation}\label{peeloff}
\frac{d}{dt}\beta_{k+1}
=i\epsilon(\lambda)\beta_{k+1}+\sum_{n=1}^{N}\Theta_{n}(\lambda)\beta_{k+1}^{n+1}\bar{\beta}_{k+1}^{n}
+D_{1}+D_{2}+D_{3}+Remainder,
\end{equation}
where the new function $\beta_{k+1}$ is defined as
\begin{equation}\label{transformation}
\beta_{k+1}:=\beta_{k}-\sum_{(m,n)\in\mathcal{A}_{k}}\frac{\Theta_{m,n}(\lambda)}{i(m-n-1)\epsilon(\lambda)}\beta_{k}^{m}\bar{\beta}_{k}^{n},
\end{equation} and we observe that
$\frac{\Theta_{m,n}(\lambda)}{i(m-n-1)\epsilon(\lambda)}$ are real
for $m,n\leq N;$ the terms $D_{n},$ $n=1,2,3,$ are given by
$$D_{1}:=\displaystyle\sum_{k+1\leq l\leq 2N+1
}\sum_{(m,n)\in\mathcal{A}_{l}}\Theta_{m,n}(\lambda)\beta_{k}^{m}\bar{\beta}_{k}^{n},$$
$$D_{2}:=\displaystyle\sum_{n=1}^{N}\Theta_{n}(\lambda)\beta_{k}^{n+1}\bar{\beta}_{k}^{n}-\displaystyle\sum_{n=1}^{N}\Theta_{n}(\lambda)\beta_{k+1}^{n+1}\bar{\beta}_{k+1}^{n},$$
$$D_{3}:=-\frac{d}{dt}\sum_{(m,n)\in\mathcal{A}_{k}}\frac{\Theta_{m,n}(\lambda)}{i(m-n-1)\epsilon(\lambda)}\beta_{k}^{m}\bar{\beta}_{k}^{n}+\sum_{(m,n)\in\mathcal{A}_{k}}\frac{(m-n)\Theta_{m,n}(\lambda)}{m-n-1}\beta_{k}^{m}\bar{\beta}_{k}^{n}.$$

By Proposition ~\ref{Co:invariant} in Appendix ~\ref{AP:Transform}
\begin{equation}\label{final}
\begin{array}{lll}
\dot{\beta}_{k+1}&=&i\epsilon(\lambda)\beta_{k+1}+\displaystyle\sum_{n=1}^{N}\Theta_{n}^{(1)}(\lambda)\beta_{k+1}^{n+1}\bar{\beta}_{k+1}^{n}+\sum_{2\leq
l\leq
2N+1}\sum_{(m,n)\in\mathcal{A}_{l}}\Theta^{(1)}_{m,n}(\lambda)\beta_{k+1}^{m}\bar{\beta}_{k+1}^{n}\\
& &+Remainder
\end{array}
\end{equation}
with $\Theta^{(1)}_{m,n}(\lambda)$ being purely imaginary if
$m,n\leq N$ and $\Theta^{(1)}_{n}(\lambda)$ are purely imaginary if
$n<N.$ We claim that $\Theta^{(1)}_{m,n}(\lambda)=0$ for $(m,n)\in
\displaystyle\cup_{l\leq k}\mathcal{A}_{l}$. This is due to the fact
that the terms $D_{n}, \ n=1,2,3,$ in Equation (~\ref{peeloff}) are
of the order $O(|\beta|^{k+1})$. This relations together with
Equation (~\ref{final}) imply
$$
\begin{array}{lll}
\dot{\beta}_{k+1}&=&i\epsilon(\lambda)\beta_{k+1}+\displaystyle\sum_{n=1}^{N}\Theta_{n}^{(1)}(\lambda)\beta^{n+1}_{k+1}\bar{\beta}_{k+1}^{n}+\displaystyle\sum_{k+1\leq l\leq 2N+1}\displaystyle\sum_{(m,n)\in\mathcal{A}_{l}}\Theta^{(1)}_{m,n}(\lambda)\beta_{k+1}^{m}\bar{\beta}_{k+1}^{n}\\
& &+Remainder
\end{array}
$$
where $\Theta_{m,n}^{(1)}(\lambda)$ are purely imaginary for
$m,n\leq N$ and $\Theta^{(1)}_{n}(\lambda)$ are purely imaginary for
$n<N.$ Thus we complete the induction steps. Taking
$\beta=\beta_{2N+1}$ we see that $\beta$ satisfies the statement (A)
of Lemma ~\ref{transformz}.

Now we prove Statement (B). By Statement (A)
\begin{equation}\label{defineBeta}
\beta=y+\sum_{N+1\leq m+n\leq
2N+1}u_{mn}(\lambda)y^{m}\bar{y}^{n}\end{equation} with
$u_{mn}(\lambda)$ being real for $m,n\leq N$. We invert this
function to get the relation
$$y=\beta-\sum_{N+1\leq
m+n\leq 2N+1}u_{mn}(\lambda)\beta^{m}\bar{\beta}^{n}+Remainder$$
where $u_{m,n}(\lambda)$ are the same as in (~\ref{defineBeta}). We
substitute the expression for $y$ in Equation (~\ref{expanlambda})
to obtain
$$\dot\lambda=\sum_{2\leq m+n\leq
2N+1}\lambda_{mn}(\lambda)\beta^{m}\bar{\beta}^{n}+Remainder$$ with
$$\lambda_{m,n}(\lambda):=\Lambda_{m,n}(\lambda)-\sum_{\begin{subarray}{lll}
m'+l_{1}=m+1\\
n'+l_{2}=n
\end{subarray}
}l_{1}u_{m',n'}\Lambda_{l_{1},l_{2}}(\lambda)-\sum_{\begin{subarray}{lll}
m'+l_{2}=m\\
n'+l_{1}=n+1
\end{subarray}
}l_{2}u_{m',n'}\bar\Lambda_{l_{1},l_{2}}(\lambda).$$ If $m,n\leq N,$
$l_{1}\not=0,$ $m'+l_{1}=m+1$, $n'+l_{2}=n$ and $m'+n'\geq N+1$ then
$l_{1},l_{2},m',n'\leq N$. Thus
$l_{1}u_{m',n'}\Lambda_{l_{1},l_{2}}(\lambda)$ in the equation above
are purely imaginary, where, recall the property of
$\Lambda_{m,n}(\lambda)$ from (~\ref{expanlambda}) if
$m'+l_{1}-1=m,\ n'+l_{2}=n\leq N$. Similarly
$l_{2}u_{m',n'}\bar\Lambda_{l_{1},l_{2}}(\lambda)$ is purely
imaginary if $m'+l_{2}=m,\ n'+l_{1}-1=n\leq N $. Therefore
$\lambda_{m,n}(\lambda)$ is purely imaginary for $m,n\leq N.$
\begin{flushright}
$\square$
\end{flushright}
\section{The Decay of $y$}\label{se:zdecay}
Let the parameter $\beta$ be the same as in Lemma ~\ref{transformz}.
Recall that $Re Y_{N}(\lambda)< 0$ by Condition (FGR) in Theorem
~\ref{maintheorem1}. We have
\begin{lemma}\label{Ydecay} for any $t\leq T$ we have
\begin{equation}\label{controlY}
|y|, |\beta|\leq
c(T_{0}+t)^{-\frac{1}{2N}}[1+T_{0}^{-\frac{1}{2N}}(Y(T)+\mathcal{R}_{1}(T))^{2}]
\end{equation}
for some constant $c.$
\end{lemma}
\begin{proof}
For any $t\geq 0$, define
$$X(t):=\displaystyle\sup_{s\leq t}(T_{0}+s)^{\frac{1}{2N}}|\beta(s)|.$$
By the relationship between $\beta$ and $y$ we have that if $X$ is
uniformly bounded in $t$, then
\begin{equation}\label{equivalent}
cY\leq X\leq \frac{1}{c}Y
\end{equation}
for some constant $c,$ where, recall the functions $Y=Y(t),
\mathcal{R}_{n}=\mathcal{R}_{n}(t),\ n=1,2,3,$ defined in Equation
(~\ref{eq:majorants}). We claim that
\begin{equation}\label{majorantX}
X\leq cX(0)[1+(Y+\mathcal{R}_{1})^{2}].
\end{equation}
Indeed, by the equation in Statement (A) of Lemma ~\ref{transformz}
we have that
\begin{equation}\label{prericatti}
\frac{1}{2}\frac{d}{dt}|\beta|^{2}=ReY_{N}(\lambda)|\beta|^{2N+2}+Re(\bar{\beta}\text{Remainder})
\end{equation}
which can be transformed into a Riccati equation
\begin{equation}\label{eq:Reccati}
\begin{array}{lll}
\frac{1}{2N}\frac{d}{dt}|\beta|^{2N}&=&ReY_{N}(\lambda)|\beta|^{4N}+|\beta|^{2N-2}Re(\bar{\beta}\text{Remainder})\\
&\leq &ReY_{N}(\lambda)|\beta|^{4N}+|\beta|^{2N-1}|Remainder|.
\end{array}
\end{equation}
By the estimate of $Remainder$ in Equation (~\ref{remainder}), the
property $Re Y_{N}(\lambda)<0$ (see Condition (FGR)) and Equations
(~\ref{equivalent}) and (~\ref{eq:Reccati}) we have Equation
(~\ref{majorantX}). This together with Equation (~\ref{equivalent})
implies Lemma ~\ref{Ydecay}.
\end{proof}
\section{Proof of the Main Theorems ~\ref{maintheorem1} and
~\ref{maintheorem2} for $d\geq 3$}\label{ProveMain} In order not to
complicate notations we construct the proof of the main Theorems
~\ref{maintheorem1} and ~\ref{maintheorem2} for $d=3$ rather than
$d\geq 3$. This proof can be easily modified to obtain the general
$d\geq 3$ cases (the only difference is that one has to deal with
$[\frac{d}{2}]+3$ derivatives, see Subsection
~\ref{Sec:Propagator}). We begin with some preliminary results. The
following lemma will be used repeatedly.
\begin{lemma}\label{mainpart}
There is a constant $\epsilon>0$ such that if
$|\lambda-\lambda_{1}|\leq \epsilon$ then there is a constant $c>0$
such that
\begin{equation}\label{eq:R2main}
\begin{array}{ccc}
& &\|\rho_{\nu}(-\Delta+1)^{2}R_{N}\|_{2}\leq
c\|\rho_{\nu}(-\Delta+1)^{2}P_{c}^{\lambda_{1}}R_{N}\|_{2},\\
& &\|R_{N}\|_{\infty}\leq c\|P_{c}^{\lambda_{1}}R_{N}\|_{\infty},\\
& &\|R_{N}\|_{\mathcal{H}^{2}}\leq
c\|P_{c}^{\lambda_{1}}R_{N}\|_{\mathcal{H}^{2}},
\end{array}
\end{equation}
\begin{equation}\label{eq:gmain}
\|\rho_{\nu}R_{2N}\|_{2}\leq
c\|\rho_{\nu}P_{c}^{\lambda_{1}}R_{2N}\|_{2}.
\end{equation}
\end{lemma}
\begin{proof}
We only prove the first three estimates, the proof of
(~\ref{eq:gmain}) is similar. First, since the vectors
$$\xi_{1}:=\left(
\begin{array}{lll}
0\\
\phi^{\lambda_{1}}
\end{array}
\right),\ \xi_{2}:=\left(
\begin{array}{lll}
\frac{d}{d\lambda}\phi^{\lambda_{1}}\\
0
\end{array}
\right),\ \xi_{3}:=\left(
\begin{array}{lll}
\xi^{\lambda_{1}}\\
0
\end{array}
\right),\ \xi_{4}:=\left(
\begin{array}{lll}
0\\
\eta^{\lambda_{1}}
\end{array}
\right)$$ span the space $Range\{1-P_{c}^{\lambda_{1}}\}$ there
exists a vector $\vec{a}=(a_{1},\cdot\cdot\cdot,a_{4})$ such that
\begin{equation}\label{Decom}
R_{N}=P_{c}^{\lambda_{1}}R_{N}+\sum_{n=1}^{4} a_{n}\xi_{n}.
\end{equation} From the equation $(1-P_{c}^{\lambda})R_{N}=0$ we derive
the equation $A\vec{a}=-\vec{b} $ where $\vec{b}$ is a $4\times 1$
vector with the components $b_{j}:=\langle
P_{c}^{\lambda_{1}}R_{N},\xi_{j}\rangle,$ and $A$ is the $4\times 4$
matrix
$$A:=
\left(
\begin{array}{cccc}
0&\langle
\phi^{\lambda},\frac{d}{d\lambda}\phi^{\lambda_{1}}\rangle&\langle
\phi^{\lambda},\xi^{\lambda_{1}}\rangle&0\\
\langle
\phi^{\lambda_{1}},\frac{d}{d\lambda}\phi^{\lambda}\rangle&0&0&\langle
\eta^{\lambda_{1}},\frac{d}{d\lambda}\phi^{\lambda_{1}}\rangle\\
0&\langle
\frac{d}{d\lambda}\phi^{\lambda_{1}},\eta^{\lambda}\rangle&\langle
\xi^{\lambda_{1}},\eta^{\lambda}\rangle&0\\
\langle \phi^{\lambda_{1}},\xi^{\lambda}\rangle &0 &0 &\langle
\eta^{\lambda_{1}},\xi^{\lambda}\rangle
\end{array}
\right).
$$ By the fact that $\lambda\in \mathcal{I}$ (the interval $\mathcal{I}$ is defined in Equation
(~\ref{Stab})) and Equation (~\ref{xieta}) we have $$\langle
\phi^{\lambda},\frac{d}{d\lambda}\phi^{\lambda}\rangle,\ \langle
\xi^{\lambda},\eta^{\lambda}\rangle\geq c>0$$ for some constant $c$
and by (~\ref{SymNo}) $$\langle
\phi^{\lambda},\xi^{\lambda}\rangle=\langle
\frac{d}{d\lambda}\phi^{\lambda},\eta^{\lambda}\rangle=0.$$

Thus if $|\lambda-\lambda_{1}|$ is small, then the matrix $A$ is
invertible and $\|A^{-1}\|\leq C$ for some constant $C$. Thus
$\vec{a}=-A^{-1}\vec{b}$ and therefore $|\vec{a}\leq c|\vec{b}|.$ By
Equation (~\ref{Decom}) and the definition of $\vec{b}$ we have
$$\|(1-P_{c}^{\lambda_{1}})R_{N}\|\leq c\|P_{c}^{\lambda_{1}}R_{N}\|$$
in the spaces $\rho_{-\nu}\mathcal{H}^{4},$ $\mathcal{H}^{2}$ and
$\mathcal{L}_{\infty}.$ Thus $$\|R_{N}\|\leq
\|P_{c}^{\lambda_{1}}R_{N}\|+\|(1-P_{c}^{\lambda_{1}})R_{N}\|\leq
(c+1)\|P_{c}^{\lambda_{1}}R_{N}\|$$ which is Equation
(~\ref{eq:R2main}).
\end{proof}
\subsection{Estimates on the Propagator}\label{Sec:Propagator} We will need the
following estimates of the evolution operator
$U(t):=e^{tL(\lambda_{1})}$ where $\lambda_{1}:=\lambda(T)$ for some
fixed $T\geq 0$, which we formulate in the general case $d\geq 3$
though we consider presently only the case $d=3:$
\begin{equation}\label{second}
\|\rho_{\nu}(-\Delta+1)^{k} U(t)P_{c}^{\lambda_{1}} h\|_{2}\leq
c(1+t)^{-\frac{d}{2}}\|\rho_{-\nu}(-\Delta+1)^{k}h\|_{2};
\end{equation}
\begin{equation}\label{zeromodeestimate}
\begin{array}{lll}
& &\|\rho_{\nu}(-\Delta+1)^{k}\displaystyle\prod(L(\lambda)-ik\epsilon(\lambda)+i0)^{-n_{k}}U(t)P_{c}^{\lambda_{1}}h\|_{2}\\
&\leq& c (1+t)^{-d/2}\|e^{\epsilon|x|}(-\Delta+1)^{k}h\|_{2}
\end{array}
\end{equation} with $\sum{n_{k}\leq
2N};$
\begin{equation}\label{third}
\|U(t)P_{c}^{\lambda_{1}} h\|_{\mathcal{L}^{\infty}}\leq
ct^{-d/2}\|h\|_{1};
\end{equation}
\begin{equation}\label{lastestimate}
\|U(t)P_{c}^{\lambda_{1}}h\|_{\infty}\leq
c(1+t)^{-d/2}(\|h\|_{\mathcal{H}^{k}}+\|h\|_{1});
\end{equation}
\begin{equation}\label{finalestimate}
\|\rho_{\nu}(-\Delta+1)^{k}U(t)P_{c}^{\lambda_{1}}h\|_{2}\leq
c(1+t)^{-d/2}(\|(-\Delta+1)^{k}h\|_{1}+\|(-\Delta+1)^{k}h\|_{2})
\end{equation}
where $\epsilon$ is any positive constant, $k:=[\frac{d}{2}]+1$ and
$\nu$ is a large positive constant depending on $N$. Estimate
(~\ref{second}) comes from the estimate
$$\|\rho_{\nu} U(t)P_{c}^{\lambda_{1}} h\|_{2}\leq
c(1+t)^{-\frac{d}{2}}\|\rho_{-\nu}h\|_{2}$$ proved in ~\cite{RSS},
and the observation that
\begin{equation}
\begin{array}{lll}
\|\rho_{\nu} (-\Delta+1)^{k}U(t)P_{c}^{\lambda_{1}} h\|_{2}&\leq&
c\|\rho_{\nu} (L(\lambda_{1})+k_{1})^{k}U(t)P_{c}^{\lambda_{1}}
h\|_{2}\\
&=&c\|\rho_{\nu} U(t)P_{c}^{\lambda_{1}} (L(\lambda_{1})+k_{1})^{k}
h\|_{2}
\end{array}
\end{equation} for some constant $k_{1},$ and $k:=[\frac{d}{2}]+1.$
Estimate (~\ref{third}) can be proved by the same technique as in
~\cite{GoSc} where a version of this estimate for the case of
self-adjoint operators is proved. Estimates (~\ref{lastestimate})
and (~\ref{finalestimate}) follow from Estimate (~\ref{third}) (the
long time part) and the estimate
$$\|U(t)P_{c}^{\lambda_{1}}h\|_{\infty}\leq c\|U(t)P_{c}^{\lambda_{1}}h\|_{\mathcal{H}_{k}}\leq
c\|h\|_{\mathcal{H}^{k}}$$ and
$$\|\rho_{\nu}(-\Delta+1)^{k}U(t)P_{c}^{\lambda_{1}}h\|_{2}\leq \|(-\Delta+1)^{k}U(t)P_{c}^{\lambda_{1}}h\|_{\infty}$$
(the short time part). Estimate (~\ref{zeromodeestimate}) comes from
Estimate (~\ref{second}) and the technique of deformation of contour
of integration from ~\cite{BuSu, Rauch, RSS}.

In the next subsections we begin estimating the majorants
$\mathcal{R}_{n},\ n=1,2,$ and $Y$ defined in Equation
(~\ref{majorant}). We write
$$\mathcal{R}_{1}=\mathcal{R}_{a}+\mathcal{R}_{b}+\mathcal{R}_{c}$$
where
\begin{equation}\label{eq:majorants}
\begin{array}{lll}
\mathcal{R}_{a}(T)&:=&\displaystyle\max_{t\leq
T}(T_{0}+t)^{\frac{N+1}{2N}}\|\rho_{\mu}R_{N}\|_{\mathcal{H}^{l}},\\
\mathcal{R}_{b}(T)&:=&\displaystyle\max_{t\leq
T}(T_{0}+t)^{\frac{N+1}{2N}}\|\rho_{\mu}R_{N}\|_{\infty},\\
\mathcal{R}_{c}(T)&:=&\displaystyle\max_{t\leq
T}(T_{0}+t)^{\frac{2N+1}{2N}}\|\rho_{\mu}R_{2N}\|_{2},
\end{array}
\end{equation} and recall the definitions of the constants $l$ and
$T_{0}$ after (~\ref{majorant}), and estimate the estimating
functions $\mathcal{R}_{a},\ \mathcal{R}_{b},\ \mathcal{R}_{c}$
separately.
\subsection{Estimate for
$\mathcal{R}_{a}$} The following proposition is the main result of
this subsection.
\begin{proposition}\label{controlM1}
$$\mathcal{R}_{a}\leq cT_{0}^{\frac{N+1}{2N}}\|\rho_{-2}R_{N}(0)\|_{2}+c(T_{0}^{-\frac{1}{2N}}Y\mathcal{R}_{a}+Y^{N+1}+Y^{2}\mathcal{R}_{a}^{2}\mathcal{R}_{2}^{2}+Y^{2}\mathcal{R}_{a}\mathcal{R}_{b}^{2}\mathcal{R}_{2}^{3}+\mathcal{R}_{b}^{3}\mathcal{R}_{2}^{4}].$$
\end{proposition}
Before proving the proposition we derive a new equation for $R_{N}.$
If we write
$L(\lambda(t))=L(\lambda_{1})+L(\lambda(t))-L(\lambda_{1})$, then
Equation (~\ref{equationg}) for $R_{N}$ takes the form
$$\frac{d}{dt}P^{\lambda_{1}}_{c}R_{N}=L(\lambda_{1})P^{\lambda_{1}}_{c}R_{N}+(\lambda-\lambda_{1}+\dot\gamma)P^{\lambda_{1}}_{c}\sigma_{3}R_{N}+\cdot\cdot\cdot$$
where $\sigma_{3}:=\left(
\begin{array}{lll}
1&0\\
0&-1
\end{array}
\right).$ The propagator generated by the operator
$L(\lambda_{1})+(\lambda-\lambda_{1}+\dot\gamma)P_{c}^{\lambda}\sigma_{3}$
is estimated using the following extension of a result from
~\cite{BuSu} whose proof we omit. Denote by $P_{+}$ and $P_{-}$ the
projection operators onto the positive and negative branches of the
essential spectrum of $L(\lambda_{1}),$ respectively. Then we have
\begin{lemma}\label{LM:ApproOp} For any function $h$ we have
$$\|\rho_{-\nu}(-\Delta+1)^{2}(P_{c}^{\lambda_{1}}\sigma_{3}-iP_{+}+iP_{-})h\|_{2}\leq c\|\rho_{\nu}(-\Delta+1)^{2}h\|_{2}$$
for any large $\nu>0$.
\end{lemma}
Equation (~\ref{equationg}) can be rewritten as
\begin{equation}\label{estimater21r22}
\begin{array}{lll}
\frac{d}{dt}P_{c}^{\lambda_{1}}R_{N}&=&L(\lambda_{1})P_{c}^{\lambda_{1}}
R_{N}+[\dot{\gamma}+\lambda-\lambda_{1}]i(P_{+}-P_{-})R_{N}\\
&
&+P_{c}^{\lambda_{1}}O_{1}R_{N}+P_{c}^{\lambda_{1}}F_{N}(y,\bar{y})+P_{c}^{\lambda_{1}}N_{N}(R_{N},y,\bar{y}),
\end{array}
\end{equation}
where $O_{1}$ is the operator defined by
\begin{equation}\label{EstO1}
O_{1}:=P_{N}(y,\bar{y})+\dot{\lambda}P_{c\lambda}+L(\lambda)-L(\lambda_{1})+\dot{\gamma}P_{c}^{\lambda}\sigma_{3}-[\dot{\gamma}+\lambda-\lambda_{1}]i(P_{+}-P_{-})
\end{equation}
and the definitions of and estimates on $P_{N}(y,\bar{y})$ and
$N_{N}(R_{N},y,\bar{y})$ are given in Theorem ~\ref{expansion}, Part
(RC). Equations (~\ref{majorant}), (~\ref{eq:majorants}),
(~\ref{convergencelambda}) and (~\ref{controlY}) imply that
\begin{equation}\label{estimateonO1}
\|\rho_{-\nu}(-\Delta+1)^{2}O_{1}R_{N}\|_{2}\leq
c(T_{0}+t)^{-\frac{N+2}{2N}}Y\mathcal{R}_{a},
\end{equation}
\begin{equation}\label{estimateonO2}
\|\rho_{-\nu}(-\Delta+1)^{2}F_{N}(y,\bar{y})\|_{2}\leq
c(T_{0}+t)^{-\frac{N+1}{2N}}Y^{N+1},
\end{equation}
\begin{equation}\label{estimateonO3}
\begin{array}{lll}
& &\|(-\Delta+1)^{2}N_{N}(R_{N},y,\bar{y})\|_{1}+\|(-\Delta+1)^{2}N_{N}(R_{N},y,\bar{y})\|_{2}\\
&\leq&
c(T_{0}+t)^{-\frac{2N+3}{2N}}(Y^{2}\mathcal{R}_{a}^{2}\mathcal{R}_{2}^{2}+Y^{2}\mathcal{R}_{a}\mathcal{R}_{b}^{2}\mathcal{R}_{2}^{3}+\mathcal{R}_{b}^{3}\mathcal{R}_{2}^{4}).
\end{array}
\end{equation}
By Equation (~\ref{estimater21r22}) and the observation that the
operators $P_{+},$ $P_{-}$ and $L(\lambda_{1})$ commute with each
other, we have
\begin{equation}\label{rewriter21r22}
\begin{array}{lll}
P_{c}^{\lambda_{1}}R_{N}
&=&e^{tL(\lambda_{1})+a(t,0)(P_{+}-P_{-})}P_{c}^{\lambda_{1}}R_{N}(0)\\
& &+\int_{0}^{t}e^{(t-s)L(\lambda_{1})+a(t,s)(P_{+}-P_{-})}P_{c}^{\lambda_{1}}[O_{1}R_{N}\\
& &+F_{N}(y,\bar{y})+N_{N}(R_{N},y,\bar{y})]ds,
\end{array}
\end{equation} with $a(t,s):=\int_{s}^{t}
i[\dot{\gamma}(k)+\lambda(k)-\lambda_{1}]dk.$ We observe that
$P_{+}P_{-}=P_{-}P_{+}=0$ and for any times $t_{1}\leq t_{2}$ the
operator
$$
e^{ a(t_{2},t_{1})(P_{+}-P_{-})}
=e^{a(t_{2},t_{1})}P_{+}+e^{-a(t_{2},t_{1})}P_{-}:
\mathcal{H}^{4}\rightarrow \mathcal{H}^{4}
$$ is uniformly bounded.
Now we prove Proposition ~\ref{controlM1}.\\ {\bf{Proof of
Proposition~\ref{controlM1}}}. By Equation (~\ref{rewriter21r22}),
Estimates (~\ref{second}) and (~\ref{finalestimate}) for $d=3$ we
have
\begin{equation}\label{estimaterinnorm2}
\begin{array}{lll}
& &\|\rho_{\nu}(-\Delta+1)^{2}P_{c}^{\lambda_{1}}R_{N}(t)\|_{2}\\
&\leq&
\|\rho_{\nu}(-\Delta+1)^{2}e^{tL(\lambda_{1})}P_{c}^{\lambda_{1}}R_{N}(0)\|_{2}\\
&
&+\|\int_{0}^{t}\rho_{\nu}(-\Delta+1)^{2}e^{(t-s)L(\lambda_{1})}P_{c}^{\lambda_{1}}[O_{1}(s)R_{N}+F_{N}(y,\bar{y})+N_{N}(R_{N},y,\bar{y})]ds\|_{2}\\
&\leq&c(1+t)^{-3/2}\|\rho_{-\nu}(-\Delta+1)^{2}R_{N}(0)\|_{2}\\
&
&+\int_{0}^{t}(1+t-s)^{-3/2}\|\rho_{-\nu}(-\Delta+1)^{2}[O_{1}R_{N}
+F_{N}(y,\bar{y})]ds\|_{2}\\
&
&+\int_{0}^{t}(1+t-s)^{-3/2}(\|(-\Delta+1)^{2}N_{N}(R_{N}(s),y,\bar{y})\|_{1}
+\|(-\Delta+1)^{2}N_{N}(R_{N}(s),y,\bar{y})\|_{2})ds.
\end{array}
\end{equation}
Therefore by Lemma ~\ref{mainpart} and Estimates
(~\ref{estimateonO1})-(~\ref{estimateonO3}) we have
$$
\begin{array}{lll}
& &\|\rho_{\nu}(-\Delta+1)^{2}R_{N}\|_{2}\\ &\leq&
c_{1}\|\rho_{\nu}(-\Delta+1)^{2}P_{c}^{\lambda_{1}}R_{N}\|_{2}\\
&\leq&c_{2}[(1+t)^{-3/2}\|\rho_{-\nu}(-\Delta+1)^{2}R_{N}(0)\|_{2}
+\int_{0}^{t}(1+t-s)^{-3/2}(T_{0}+s)^{-\frac{N+1}{2N}}ds\\&
&\times(T_{0}^{-\frac{1}{2N}}Y\mathcal{R}_{a}+Y^{N+1}+Y^{2}\mathcal{R}_{a}^{2}\mathcal{R}_{2}^{2}+Y^{2}\mathcal{R}_{a}\mathcal{R}_{b}^{2}\mathcal{R}_{2}^{3}+\mathcal{R}_{b}^{3}\mathcal{R}_{2}^{4}].
\end{array}
$$
Using the estimate
$$\int_{0}^{t}(1+t-s)^{-3/2}(T_{0}+s)^{-\frac{N+1}{2N}}ds\leq
c(T_{0}+t)^{-\frac{N+1}{2N}}$$ we obtain
$$
\begin{array}{lll}
\|\rho_{\nu}(-\Delta+1)^{2}R_{N}\|_{2}&\leq&
c(T_{0}+t)^{-\frac{N+1}{2N}}[T_{0}^{\frac{N+1}{2N}}\|\rho_{-2}R_{N}(0)\|_{2}+T_{0}^{-\frac{1}{2N}}Y\mathcal{R}_{a}+Y^{N+1}\\
& &+Y^{2}\mathcal{R}_{a}^{2}\mathcal{R}_{2}^{2}
+Y^{2}\mathcal{R}_{a}\mathcal{R}_{b}^{2}\mathcal{R}_{2}^{3}+\mathcal{R}_{b}^{3}\mathcal{R}_{2}^{4}].
\end{array}
$$
This and the definition of $\mathcal{R}_{a}$ (in Equation
(~\ref{eq:majorants})) imply Proposition ~\ref{controlM1}.
\begin{flushright}
$\square$
\end{flushright}
\subsection{Estimate for $\mathcal{R}_{b}$} The following proposition is the main result
of this subsection.
\begin{proposition}\label{controlM2}
$$\mathcal{R}_{b}\leq c[T_{0}^{\frac{N+1}{2N}}\|R_{N}(0)\|_{1}+T_{0}^{\frac{N+1}{2N}}\|R_{N}(0)\|_{\mathcal{H}^{2}}+T_{0}^{-\frac{1}{2N}}Y\mathcal{R}_{a}+Y^{2}\mathcal{R}_{a}^{2}\mathcal{R}_{b}^{4}+\mathcal{R}_{b}^{5}\mathcal{R}_{2}^{2}].$$
\end{proposition}
\begin{proof}
By Estimate (~\ref{lastestimate}) in $d=3$, Lemma ~\ref{mainpart}
and Equation (~\ref{estimater21r22}) we have that
\begin{equation}
\begin{array}{lll}
& &\|R_{N}(t)\|_{\infty} \\
&\leq&c\|P_{c}^{\lambda_{1}}R_{N}(t)\|_{\infty}\\
&\leq& c\|e^{tL(\lambda_{1})}P_{c}^{\lambda_{1}}R_{N}(0)\|_{\infty}
+\int_{0}^{t}\|e^{(t-s)L(\lambda_{1})}P_{c}^{\lambda_{1}}[O_{1}(s)R_{N}+F_{N}(y,\bar{y})+N_{N}(R_{N},y,\bar{y})]\|_{\infty}ds\\
&\leq&c(1+t)^{-3/2}(\|R_{N}(0)\|_{1}+\|R_{N}(0)\|_{\mathcal{H}^{2}})\\
& &+c\int_{0}^{t}(1+t-s)^{-3/2}[\|O_{1}(s)R_{N}+F_{N}(y,\bar{y})\|_{1}+\|O_{1}(s)R_{N}+F_{N}(y,\bar{y})\|_{\mathcal{H}^{2}}]ds\\
&
&+c\int_{0}^{t}(1+t-s)^{-3/2}(\|N_{N}(R_{N},y,\bar{y})\|_{1}+\|N_{N}(R_{N},y,\bar{y})\|_{\mathcal{H}^{2}})ds.
\end{array}
\end{equation}
By the properties of $O_{1}$ (Equation (~\ref{EstO1})) and $F_{N}$
(Equation (~\ref{equationg})) we have
$$\|O_{1}(s)R_{N}+F_{N}(y,\bar{y})\|_{1}+\|O_{1}(s)R_{N}+F_{N}(y,\bar{y})\|_{\mathcal{H}^{2}}\leq c(T_{0}+t)^{-\frac{N+1}{2N}}Y(T)\mathcal{R}_{a}(T).$$
By Equation (~\ref{remainderestiamte2})
$$\|N_{N}(R_{N},y,\bar{y})\|_{1}+\|N_{N}(R_{N},y,\bar{y})\|_{\mathcal{H}^{2}}\leq c (T_{0}+t)^{-\frac{2N+3}{2N}}[Y^{2}\mathcal{R}_{a}^{2}\mathcal{R}_{b}^{4}+\mathcal{R}_{b}^{5}\mathcal{R}_{2}^{2}].$$
Hence
$$
\begin{array}{lll}
\|R_{N}(t)\|_{\infty} &\leq&
c(T_{0}+t)^{-\frac{N+1}{2N}}[T_{0}^{\frac{N+1}{2N}}\|R_{N}(0)\|_{1}\\
&
&+T_{0}^{\frac{N+1}{2N}}\|R_{N}(0)\|_{\mathcal{H}^{2}}+T_{0}^{-\frac{1}{2N}}Y(t)\mathcal{R}_{a}(t)+Y\mathcal{R}_{a}^{2}\mathcal{R}_{b}^{4}(t)+\mathcal{R}_{b}^{5}\mathcal{R}_{2}^{2}(t)].
\end{array}
$$
This estimate and the definition of $\mathcal{R}_{b}$ yield the
proposition.
\end{proof}
\subsection{Estimate for $\mathcal{R}_{c}$}
The following is the main result of this subsection
\begin{proposition}\label{controlM3} Let the constant $\nu$ the same as in
(~\ref{second})-(~\ref{zeromodeestimate}) with $d=3$. Then
\begin{equation}\label{EstR3}
\mathcal{R}_{c}\leq
c[T_{0}^{\frac{2N+1}{2N}}\|\rho_{-\nu}R_{N}(0)\|_{2}+T_{0}^{\frac{2N+1}{2N}}|y|^{N+1}(0)]+c(T_{0}^{-\frac{1}{2N}}Y\mathcal{R}_{c}+Y^{2N+1}+Y^{2}\mathcal{R}_{a}^{2}\mathcal{R}_{b}^{4}+\mathcal{R}_{b}^{5}\mathcal{R}_{2}^{2}).
\end{equation}
\end{proposition}
\begin{proof}
By the same techniques as we used in deriving Equation
(~\ref{estimater21r22}) we have the following equation
\begin{equation}\label{estimateg1g2}
\begin{array}{lll}
\frac{d}{dt}P_{c}^{\lambda_{1}}R_{2N}&=&L(\lambda_{1})P_{c}^{\lambda_{1}}R_{2N}+(\dot\gamma+\lambda-\lambda_{1})i(P_{+}-P_{-})R_{2N}\\
&
&+P(y,\bar{y})R_{2N}+P_{c}^{\lambda_{1}}F_{2N}(y,\bar{y})+P_{c}^{\lambda_{1}}N_{N}(R_{N},y,\bar{y}),
\end{array}
\end{equation} where the operator
$P(y,\bar{y})$ is defined as
$$P(y,\bar{y}):=P_{c}^{\lambda_{1}}P_{2N}(y,\bar{y})-(\dot{\gamma}+\lambda-\lambda_{1})i(P_{+}-P_{-})+P^{\lambda_{1}}_{c}(L(\lambda)-L(\lambda_{1})),$$
and the terms $F_{2N}(y,\bar{y})$, $P_{2N}(z,\bar{z})$ and
$N_{N}(R_{N},y,\bar{y})$ are defined in Theorem ~\ref{expansion}.

Rewriting Equation (~\ref{estimateg1g2}) in the integral form using
the Duhamel principle and using Lemma ~\ref{mainpart} we obtain
\begin{equation}\label{formulag}
\begin{array}{lll}
\|\rho_{\nu}R_{2N}(t)\|_{2}&\leq&c\|\rho_{\nu}P_{c}^{\lambda_{1}}R_{2N}(t)\|_{2}\\
&\leq&c\|\rho_{\nu}e^{tL(\lambda_{1})}P_{c}^{\lambda_{1}}R_{2N}(0)\|_{2}
+c\int_{0}^{t}\|\rho_{\nu}e^{(t-s)L(\lambda_{1})}\\
&
&\times[P(y,\bar{y})R_{2N}+P_{c}^{\lambda_{1}}F_{2N}(y,\bar{y})+N_{N}(R_{N},y,\bar{y})]\|_{2}ds.
\end{array}
\end{equation}
We claim that
\begin{equation}\label{estimateg1g20}
 \|\rho_{\nu}e^{tL(\lambda_{1})}P_{c}^{\lambda_{1}}R_{2N}(0)\|_{2}\leq c(1+t)^{-3/2}(\|\rho_{-\nu}R_{N}(0)\|_{2}+|z(0)|^{N+1}).
\end{equation} Indeed, by Equations
(~\ref{expansionR1}) we have that
$$R_{2N}=R_{N}+\sum_{
N+1\leq m+n\leq 2N }R_{m,n}(\lambda)y^{m}\bar{y}^{n}.
$$
Therefore, displaying the time-dependent of $R_{k}$, $\lambda$ and
$y,$
$$
\begin{array}{lll}
& &\|\rho_{\nu}e^{tL(\lambda_{1})}P_{c}^{\lambda_{1}}R_{2N}(0)\|_{2}\\
&\leq&
\|\rho_{\nu}e^{tL(\lambda_{1})}P_{c}^{\lambda_{1}}R_{N}(0)\|_{2}+\displaystyle\sum_{N+1\leq
m+n\leq
2N}|y(0)|^{m+n}\|\rho_{\nu}e^{tL(\lambda_{1})}P_{c}^{\lambda_{1}}R_{m,n}\|_{2}
\end{array}
$$
By Property (RA) of $R_{m,n}(\lambda)$ given in Theorem
~\ref{expansion} and by Estimates (~\ref{second}) and
(~\ref{zeromodeestimate}) with $d=3$ we have that
$$\|\rho_{\nu}e^{tL(\lambda_{1})}P_{c}^{\lambda_{1}}R_{m,n}\|_{2}\leq c(1+t)^{-3/2}.$$

For the second term of the right hand side of Equation
(~\ref{formulag}), we have
\begin{equation}\label{secondterm}
\begin{array}{lll}
&
&\int_{0}^{t}\|\rho_{\nu}e^{(t-s)L(\lambda_{1})}[P(y,\bar{y})R_{2N}+P_{c}^{\lambda_{1}}F_{2N}(y,\bar{y})+N_{N}(R_{N},y,\bar{y})]\|_{2}ds\\
&\leq&\int_{0}^{t}(1+t-s)^{-3/2}(\|\rho_{-\nu}P(y,\bar{y})R_{2N}\|_{2}+\|N_{N}(R_{N},y,\bar{y})\|_{1}+\|N_{N}(R_{N},y,\bar{y})\|_{2})ds\\
&
&+\int_{0}^{t}\|\rho_{\nu}e^{(t-s)L(\lambda_{1})}P_{c}^{\lambda_{1}}F_{2N}(y,\bar{y})\|_{2}ds.
\end{array}
\end{equation}
For the terms on the right hand side of Equation (~\ref{secondterm})
we have the following estimates:
\begin{enumerate}
\item[(A)] By the definition of $F_{2N}(y,\bar{y})$ in Equation
(~\ref{equationg}) and Estimate (~\ref{zeromodeestimate}) with $d=3$
we have that
$$
\begin{array}{lll}
&
&|\int_{0}^{t}\|\rho_{\nu}e^{(t-s)L(\lambda_{1})}P_{c}^{\lambda_{1}}F_{2N}(y,\bar{y})\|_{2}ds|\\&\leq&
c_{1}\int_{0}^{t}(1+t-s)^{-3/2}(T_{0}+s)^{-\frac{2N+1}{2N}}ds Y^{2N+1}\\
&\leq&c_{2}(T_{0}+t)^{-\frac{2N+1}{2N}}Y^{2N+1}.
\end{array}
$$
\item[(B)] By Estimates (~\ref{estimateonO1})-(~\ref{estimateonO3}) we have
$$
\|N_{N}(R_{N}(s),y(s))\|_{1}+\|N_{N}(R_{N}(s),y(s))\|_{2}\\
\leq
c(T_{0}+s)^{-\frac{2N+1}{2N}}[Y^{2}\mathcal{R}_{a}^{2}\mathcal{R}_{b}^{4}+\mathcal{R}_{b}^{5}\mathcal{R}_{2}^{2}].
$$
\item[(C)] By the definition of $P(y,\bar{y})$ and the estimate of $P_{2N}(y,\bar{y})$ after Equation (~\ref{equationg})
$$
\begin{array}{lll}
\|\rho_{-\nu}P(y(s),\bar{y}(s))R_{2N}(s)\|_{2}&\leq&
c|y|\|\rho_{\nu}R_{2N}(s)\|_{2}\\
&\leq& c(T_{0}+s)^{-\frac{2N+2}{2N}}Y\mathcal{R}_{c}.
\end{array}
$$
\end{enumerate}
Collecting the estimates above we find $$
\begin{array}{lll}
& &\|\rho_{\nu}R_{2N}\|_{2}\\
&\leq&c(1+t)^{-3/2}[\|\rho_{-\nu}R_{N}(0)\|_{2}+|z|^{N+1}(0)]+c\int_{0}^{t}(1+t-s)^{-3/2}(T_{0}+s)^{-\frac{2N+1}{2N}}ds\\
&
&\times(T_{0}^{-\frac{1}{2N}}Y\mathcal{R}_{c}+Y^{2N+1}+Y^{2}\mathcal{R}_{a}^{2}\mathcal{R}_{b}^{4}+\mathcal{R}_{b}^{5}\mathcal{R}_{2}^{2})\\
&\leq
&c(T_{0}+t)^{-\frac{2N+1}{2N}}[T_{0}^{\frac{2N+1}{2N}}\|\rho_{-\nu}R_{N}(0)\|_{2}+T_{0}^{\frac{2N+1}{2N}}|y|^{N+1}(0)+T_{0}^{-\frac{1}{2N}}Y\mathcal{R}_{c}+Y^{2N+1}\\
&
&+Y^{2}\mathcal{R}_{a}^{2}\mathcal{R}_{b}^{4}+\mathcal{R}_{b}^{5}\mathcal{R}_{2}^{2}].
\end{array}
$$
This and the definition of $\mathcal{R}_{c}$ yield (~\ref{EstR3}).
\end{proof}
\subsection{Estimate for $\mathcal{R}_{2}$}
\begin{proposition}\label{controlR4}
\begin{equation}\label{EstR4}
\mathcal{R}_{2}^{2}\leq
\|R_{N}(0)\|^{2}_{\mathcal{H}^{4}}+c[\mathcal{R}_{a}^{2}+Y^{2}\mathcal{R}_{a}^{2}+Y^{2}\mathcal{R}_{a}^{2}\mathcal{R}_{2}+\mathcal{R}_{2}^{5}\mathcal{R}_{b}^{3}+Y^{N+1}\mathcal{R}_{a}].
\end{equation}
\end{proposition}
\begin{proof}
By Equation (~\ref{eq:R1R22}), we have
$$
\begin{array}{lll}
& &\frac{d}{dt}\langle
(-\Delta+1)^{2}R_{N},(-\Delta+1)^{2}R_{N}\rangle\\
&=&\langle
(-\Delta+1)^{2}\frac{d}{dt}R_{N},(-\Delta+1)^{2}R_{N}\rangle+\langle(-\Delta+1)^{2}R_{N},(-\Delta+1)^{2}\frac{d}{dt}R_{N}\rangle\\
&=&\displaystyle\sum_{n=1}^{4}K_{n}
\end{array}
$$ with $$K_{1}:=\langle
(-\Delta+1)^{2}(L(\lambda)+\dot\gamma
J)R_{N},(-\Delta+1)^{2}R_{N}\rangle +\langle (-\Delta+1)^{2}R_{N},
(-\Delta+1)^{2}(L(\lambda)+\dot\gamma J)R_{N}\rangle;$$
$$K_{2}:=\dot\lambda\langle
(-\Delta+1)^{2}P_{c\lambda}R_{N},(-\Delta+1)^{2}R_{N}\rangle+\dot\lambda\langle(-\Delta+1)^{2}R_{N},(-\Delta+1)^{2}P_{c\lambda}R_{N}\rangle;$$
$$K_{3}:=\langle
(-\Delta+1)^{2}N_{N}(R_{N},y,\bar{y}),(-\Delta+1)^{2}R_{N}\rangle+\langle
(-\Delta+1)^{2}R_{N},(-\Delta+1)^{2}N_{N}(R_{N},y,\bar{y})\rangle;$$
$$K_{4}:=\langle (-\Delta+1)^{2}F_{N}(y,\bar{y}),
(-\Delta+1)^{2}R_{N}\rangle+\langle
(-\Delta+1)^{2}R_{N},(-\Delta+1)^{2}F_{N}(y,\bar{y})\rangle.$$
Recall the definition of $R_{n},\ n=1,2,3$ and $Y$ in
(~\ref{eq:majorants}).

Recall the definition of the operator $L(\lambda)$ in
(~\ref{operaL}) and use the fact that $J^{*}=-J$ to obtain
$$
|K_{1}| \leq  c\|\rho_{\nu}R_{N}\|^{2}_{\mathcal{H}^{4}}\leq
c(T_{0}+t)^{-\frac{2N+2}{2N}}\mathcal{R}_{a}^{2}.
$$
By observing that $|\dot\lambda|=O(|y|^{2})$ we have that
$$|K_{2}|\leq c|y|^{2}\|\rho_{\nu}R_{N}\|^{2}_{\mathcal{H}^{4}}\leq c(T_{0}+t)^{-\frac{2N+3}{2N}}Y^{2}(t)\mathcal{R}_{a}^{2}(t).$$
Moreover by the properties of $N_{N}(R_{N},y,\bar{y})$ in
(~\ref{remainderestimate}) we have $$|K_{3}|\leq
c(T_{0}+t)^{-\frac{2N+2}{2N}}[Y^{2}\mathcal{R}_{a}^{2}\mathcal{R}_{2}(t)+\mathcal{R}_{2}^{5}\mathcal{R}_{b}^{3}(t)].$$
By the property of $F_{N}(y,\bar{y})$ in (~\ref{equationg}) we have
$$|K_{4}|\leq c|y|^{N+1}\|\rho_{\nu}R_{N}\|_{\mathcal{H}^{4}}\leq c(T_{0}+t)^{-\frac{2N+2}{2N}}Y^{N+1}\mathcal{R}_{a}.$$

Collecting all the estimates above we have
$$
\begin{array}{lll}
& &|\frac{d}{dt}\langle
(-\Delta+1)^{2}R_{N},(-\Delta+1)^{2}R_{N}\rangle|\\
&\leq
&c(T_{0}+t)^{-\frac{2N+2}{2N}}[\mathcal{R}_{a}^{2}(t)+Y^{2}(t)\mathcal{R}_{a}^{2}(t)+Y^{2}\mathcal{R}_{a}^{2}\mathcal{R}_{2}(t)+\mathcal{R}_{2}^{5}\mathcal{R}_{b}^{3}(t)+Y^{N+1}\mathcal{R}_{a}]
\end{array}
$$
which implies that $$
\begin{array}{lll}
\|R_{N}(t)\|^{2}_{\mathcal{H}^{4}}&\leq&
\|R_{N}(0)\|^{2}_{\mathcal{H}^{4}}+c[\mathcal{R}_{a}^{2}(t)\\
&
&+Y^{2}(t)\mathcal{R}_{a}^{2}(t)+Y^{2}\mathcal{R}_{a}^{2}\mathcal{R}_{2}(t)+\mathcal{R}_{2}^{5}\mathcal{R}_{b}^{3}(t)+Y^{N+1}\mathcal{R}_{a}(t)].
\end{array}
$$
This and the definition of $\mathcal{R}_{2}$ implies (~\ref{EstR4}).
\end{proof}
\subsection{Proof of Main Theorems ~\ref{maintheorem1} and
~\ref{maintheorem2}}\label{subsec:proofsubmaintheorem} Define
$M(T):=\displaystyle\sum_{n=1}^{4}\mathcal{R}_{n}(T)$ and
\begin{equation}\label{defineS}
S:=T_{0}^{\frac{2N+1}{2N}}(\|R_{N}(0)\|_{\mathcal{H}^{4}}+\|\rho_{-\nu}R_{N}(0)\|_{2}+\|R_{N}(0)\|_{1}),
\end{equation} where, recall the definition of $T_{0}$ after
(~\ref{majorant}). If $M(0)$ is sufficiently small and $Y(0)$ is
bounded, then by Propositions ~\ref{controlM1}, ~\ref{controlM2},
~\ref{controlM3}, ~\ref{controlR4} and Equation (~\ref{controlY}) we
obtain $M(T)+Y(T)\leq \mu(S)S$, where $\mu$ is a bounded function
for $S$ small. Thus we proved that if $S$ and $M(0)$ are small, then
\begin{equation}\label{equa619a}
\|\rho_{\nu}R_{N}\|_{2},\ \|R_{N}\|_{\infty}\leq
c(T_{0}+t)^{-\frac{N+1}{2N}},\ |y(t)|\leq c(T_{0}+t)^{-\frac{1}{2N}}
\end{equation}
for some constant $c.$

To complete the proof of Theorems ~\ref{maintheorem1} and
~\ref{maintheorem2} it suffices to show that
\begin{equation}\label{Ini2}
T_{0}^{\frac{2N+1}{2N}}(\|\vec{R}(0)\|_{\mathcal{H}^{4}}+\|\rho_{-\nu}\vec{R}(0)\|_{2})
\end{equation}
being small implies that $S,$ defined in Equation (~\ref{defineS}),
is small.

By Equation (~\ref{expansionR1}) we have
\begin{equation}\label{controlM}
\begin{array}{lll}
S&\leq&
cT_{0}^{\frac{2N+1}{2N}}[\|\vec{R}(0)\|_{\mathcal{H}^{4}}+\|\rho_{-\nu}\vec{R}(0)\|_{2}+\|\vec{R}(0)\|_{1}+|y^{2}(0)|]\\
&\leq
&cT_{0}^{\frac{2N+1}{2N}}[\|\vec{R}(0)\|_{\mathcal{H}^{4}}+\|\rho_{-\nu}\vec{R}(0)\|_{2}+|y^{2}(0)|]
\end{array}
\end{equation} for some constant $c>0$.
Estimate (~\ref{controlM}) implies that if (~\ref{Ini2}) is small,
then $S$ is small, and therefore Equation (~\ref{equa619a}) holds.
By Equations (~\ref{expansionR1}) and (~\ref{equa619a}) we have
$$\|\rho_{\nu}\vec{R}(t)\|\leq c(T_{0}+t)^{-\frac{1}{N}}\ \text{and}\ |y(t)|\leq
c(T_{0}+t)^{-\frac{1}{2N}}.$$ Since (~\ref{Ini2}) is small by the
condition (~\ref{InitCond}) on the datum, this together with the
relationship $|z|=|y|+O(|y|^{2}),$ yields Statements (A) and (B) of
Theorem ~\ref{maintheorem1}.

Statement (A) of Theorem ~\ref{maintheorem2} follows from
Propositions ~\ref{Pro:lamCon}, ~\ref{controlM1},
~\ref{controlM2}-~\ref{controlR4} by taking $T\geq
t\rightarrow\infty.$ Statement (B) is proved in Lemma
~\ref{transformz}.
\begin{flushright}
$\square$
\end{flushright}
\section{Proof of Theorems ~\ref{maintheorem1} and ~\ref{maintheorem2} for $d=1$}\label{Dimen1} In this section we sketch the proof of Theorems
~\ref{maintheorem1} and ~\ref{maintheorem2} for the dimension 1.
Most of the steps of the proof are almost the same to those of the
case of $d\geq 3,$ hence we concentrate on the parts which are
different, namely, the rate of decay of $\|R_{N}\|_{\infty}$ where
$R_{N}$ is the remainder in the expansion of $R$ in
(~\ref{expansionR1}). Recall that $\rho_{\nu}:=(1+|x|)^{-\nu}.$
\begin{theorem}\label{expansion1D}
Let $d=1$. Then the nonlinearity $N_{N}(R_{N},y,\bar{y})$ in
(~\ref{equationg}) satisfies, in addition, the estimate
\begin{equation}\label{RemainderEst}
\begin{array}{lll}
& &\|\rho_{-2}N_{N}(R_{N},y,\bar{y})\|_{1}+\|N_{N}(R_{N},y,\bar{y})\|_{2}\\
&\leq&
c|y|^{2}\|\rho_{\nu_{1}}R_{N}\|_{2}^{2}\|R_{N}\|_{\infty}^{6N-1}+c\|(1+|x|)
R_{N}\|_{2}^{2}\|R_{N}\|_{\infty}^{6N+1}
\end{array}
\end{equation} for $\nu_{1}>7/2$ (see (~\ref{zeromodeestimate1})
below).
\end{theorem}
Now we prove the main theorems ~\ref{maintheorem1} and
~\ref{maintheorem2} for $d=1$. We use Theorem ~\ref{expansion1D} and
Equation (~\ref{formulag}) to estimate $R_{N}$ and $R_{2N}.$ On the
first sight we need estimates of the propagator generated by the
time-dependent operator $L(\lambda(t))$. As in the $d\geq 3$ case,
we use instead the estimates on the propagator
$U(t):=e^{tL(\lambda_{1})}$ where $\lambda_{1}:=\lambda(T)$ for some
large fixed constant $T.$ We have for $d=1$
\begin{equation}\label{second1}
\|\rho_{\nu_{1}} U(t)P_{c} h\|_{2}\leq
c(1+t)^{-\frac{3}{2}}\|\rho_{-2}h\|_{2};
\end{equation}
\begin{equation}\label{zeromodeestimate1}
\|\rho_{\nu}\prod
(L(\lambda)-ik_{n}\epsilon(\lambda)+i0)^{-n_{k_{n}}}U(t)P_{c}h\|_{2}\leq
c (1+t)^{-3/2}\|e^{-\epsilon|x|}h\|_{2}
\end{equation} with $\sum{n_{k_{n}}\leq
2N};$
\begin{equation}\label{fourth1}
\|\rho_{\nu_{1}}U(t)P_{c}h\|_{2}\leq
c(1+t)^{-3/2}(\|\rho_{-2}h\|_{1}+\|h\|_{2});
\end{equation}
\begin{equation}\label{third1}
\|U(t)P_{c} h\|_{\mathcal{L}^{\infty}}\leq
ct^{-1/2}\|\rho_{-2}h\|_{2};
\end{equation}
\begin{equation}
\|U(t)P_{c} h\|_{\mathcal{L}^{\infty}}\leq
ct^{-1/2}(\|\rho_{-2}h\|_{1}+\|h\|_{2});
\end{equation}
\begin{equation}\label{first1}
\|U(t)P_{c} h\|_{\mathcal{L}^{\infty}}\leq
c(1+t)^{-\frac{1}{2}}\|\rho_{-2}h\|_{\mathcal{H}^{1}};
\end{equation}
where $\epsilon>0$, $\nu_{1}>7/2$, and $\nu$ is a large constant
depending on $N$. Estimates (~\ref{second1})
(~\ref{fourth1})-(~\ref{first1}) were proved in ~\cite{Buslaev,
BuSu, GS1, Rauch}. To prove (~\ref{zeromodeestimate1}) we use the
technique of deformation of the contour as in the proof of
(~\ref{zeromodeestimate}) and ~\cite{BuSu}. After fixing
$L(\lambda(t))$ to be $L(\lambda_{1})$ we have the equation
$$\frac{d}{dt}P^{\lambda}_{c}R_{N}=L(\lambda_{1})P^{\lambda}_{c}R_{N}+(\lambda-\lambda_{1}+\dot\gamma)P^{\lambda}_{c}\sigma_{3}R_{N}+\cdot\cdot\cdot.$$
To estimate the propagator
$e^{t[L(\lambda_{1})+(\lambda-\lambda_{1}+\dot\gamma)P_{c}\sigma_{3}]}$
we use the following lemma similar to Lemma ~\ref{LM:ApproOp} (cf
~\cite{BuSu}), whose proof we omit.
\begin{lemma} For any function $h$ we have
$$\|(1+x^{2})(P_{c}^{\lambda}\sigma_{3}-iP_{+}+iP_{-})h\|_{2}\leq c\|\rho_{\nu}h\|_{2}$$
for any $\nu>0$.
\end{lemma}
Equation (~\ref{eq:R1R22}) can be rewritten as
\begin{equation}\label{estimater21r221}
\begin{array}{lll}
\frac{d}{dt}P_{c}^{\lambda}R_{N}&=&L(\lambda_{1})P_{c}^{\lambda}
R_{N}+[\dot{\gamma}+\lambda-\lambda_{1}]i(P_{+}-P_{-})R_{N}\\
&
&+P_{c}^{\lambda}O_{1}R_{N}+P_{c}^{\lambda}F_{N}(y,\bar{y})+P_{c}^{\lambda}N_{N}(R_{N},y,\bar{y}),
\end{array}
\end{equation}
where, recall the definitions of and estimates on $F_{N}(y,\bar{y})$
and $N_{N}(R_{N},y, \bar{y})$ given in Theorem ~\ref{expansion} and
Equation (~\ref{RemainderEst}), and $O_{1}$ is the operator defined
by
$$O_{1}:=A_{2}(z,\bar{z})+\dot{\lambda}P_{c\lambda}+L(\lambda)-L(\lambda_{1})+\dot{\gamma}P_{c}^{\lambda}\sigma_{3}-[\dot{\gamma}+\lambda-\lambda_{1}]i(P_{+}-P_{-}).$$

Note that for $d=1,$ $\|R_{N}\|_{\infty}$ has a slower decay rate.
Hence we used different estimating functions than those used in
Theorem ~\ref{THM:formulascalar}.
We replace the latter functions by the following estimating
functions
\begin{equation}\label{majorant1}
\begin{array}{lll}
\mathcal{R}_{a}(T):=\displaystyle\max_{t\leq
T}(T_{0}+t)^{\frac{N+1}{2N}}\|\rho_{\nu_{1}}R_{N}\|_{2}, & &
\mathcal{R}_{b}(T):=\displaystyle\max_{t\leq
T}(T_{0}+t)^{\frac{1}{2N}}\|R_{N}\|_{\infty},\\
\mathcal{R}_{c}(T):=\displaystyle\max_{t\leq
T}(T_{0}+t)^{\frac{2N+1}{2N}}\|\rho_{\nu_{2}}R_{2N}(t)\|_{2} & &
\end{array}
\end{equation}
with $\nu_{2}>3.5$. The estimating function $Y(t)$ stays the same.
(We use the same symbols since the estimating functions,
$\mathcal{R}_{n},\ n=1,2,a,b,c,$ defined in (~\ref{majorant}) and
(~\ref{eq:majorants}), are not used in this section.)

The next lemma is proved similarly to Equations
(~\ref{estimateonO1}), (~\ref{estimateonO2}) in the $d\geq 3$ case.
\begin{lemma}\label{estimateonO11}
$$\|\rho_{-2}O_{1}R_{N}\|_{2}\leq
c(T_{0}+t)^{-\frac{N+2}{2N}}Y(T)\mathcal{R}_{a}(T),$$
$$\|\rho_{-2}F_{2}(y,\bar{y})\|_{2}\leq c(T_{0}+t)^{-\frac{N+1}{2N}}Y^{N+1}.$$
\end{lemma}
\subsection{Estimates of $\mathcal{R}_{n},\ n=a,b,c, \ \lambda(t),\ y(t)$}
In this subsection we will estimate the functions $\mathcal{R}_{n}$
which are defined in Equation (~\ref{majorant1}), $\lambda(t)$ and
$y(t)$.
\begin{proposition}\label{generalestimate}
Estimates (~\ref{convergencelambda}) and (~\ref{controlY}) on
$|\lambda(t)-\lambda(T)|$ and $|y(t)|$ hold in the case $d=1$ also.
Moreover
$$
\begin{array}{lll}
\mathcal{R}_{a}&\leq&
cT_{0}^{\frac{N+1}{2N}}\|\rho_{-2}R_{N}(0)\|_{2}\\
&
&+c(T_{0}^{-\frac{1}{2N}}Y\mathcal{R}_{a}+Y^{N+1}+Y^{2}\mathcal{R}_{a}\mathcal{R}_{b}^{6N-1}+\mathcal{R}_{b}^{6N-1}+\mathcal{R}_{b}^{6N}),
\end{array}
$$
$$
\begin{array}{lll}
\mathcal{R}_{c}&\leq&
cT_{0}^{\frac{2N+1}{2N}}[\|\rho_{-2}R_{N}(0)\|_{2}+|y|^{N+1}(0)]\\
&
&+c(T_{0}^{-\frac{1}{2N}}Y\mathcal{R}_{c}+Y^{2N+1}+Y^{2}\mathcal{R}_{a}\mathcal{R}_{b}^{6N-1}+\mathcal{R}_{b}^{6N-1}+\mathcal{R}_{b}^{6N}).
\end{array}
$$
$$
\begin{array}{lll}
\mathcal{R}_{b}&\leq&
cT_{0}^{\frac{1}{2N}}\|\rho_{-2}R_{N}(0)\|_{\mathcal{H}^{1}}\\
&
&+c(T_{0}^{-\frac{1}{2N}}Y\mathcal{R}_{a}+Y^{N+1}+Y^{2}\mathcal{R}_{a}^{2}\mathcal{R}_{b}^{6N-3}+Y^{2}\mathcal{R}_{a}\mathcal{R}_{b}^{6N-2}+\mathcal{R}_{b}^{6N}+\mathcal{R}_{b}^{6N-1}).
\end{array}
$$
\end{proposition}
\begin{proof}
The estimates of $y(t)$, $|\lambda(t)-\lambda(T)|,$
$\mathcal{R}_{a}$ and $\mathcal{R}_{c}$ are almost the same to the
$d\geq 3$ case. Therefore we focus on the estimate of
$\mathcal{R}_{b}$ which is different. (It is in this estimate where
the condition (fC) for $d=1$ is used.)

By Lemma ~\ref{mainpart}, an integral form of Equation
(~\ref{estimater21r221}) and Equations
(~\ref{third1})-(~\ref{first1}) we have that
\begin{equation}
\begin{array}{lll}
\|R_{N}(t)\|_{\infty} &\leq&c\|P_{c}^{\lambda}R_{N}(t)\|_{\infty}\\
&\leq&
c\|e^{tL(\lambda_{1})}P_{c}^{\lambda}R_{N}(0)\|_{\infty}\\
& &+\int_{0}^{t}\|e^{(t-s)L(\lambda_{1})}P_{c}^{\lambda}[O_{1}(s)R_{N}+F_{2}(y,\bar{y})+N_{N}(R_{N},y,\bar{y})]\|ds\\
&\leq&c(1+t)^{-1/2}\|\rho_{-2}R_{N}(0)\|_{\mathcal{H}^{1}}\\
& &+\int_{0}^{t}(t-s)^{-1/2}\|\rho_{-2}[O_{1}(s)R_{N}+F_{2}(y,\bar{y})]\|_{2}ds\\
&
&+\int_{0}^{t}(t-s)^{-1/2}(\|\rho_{-2}N_{N}(R_{N},y,\bar{y})\|_{1}+\|N_{N}(R_{N},y,\bar{y})\|_{2})ds.
\end{array}
\end{equation}
Recalling the estimates in Lemma ~\ref{estimateonO11}, we obtain
$$
\begin{array}{lll}
\|
R_{N}\|_{\infty}&\leq&c_{1}[(1+t)^{-1/2}\|\rho_{-2}R_{N}(0)\|_{\mathcal{H}^{1}}+\int_{0}^{t}(t-s)^{-1/2}(T_{0}+s)^{-\frac{N+1}{2N}}ds\\
&\times&
(T_{0}^{-\frac{1}{2N}}Y\mathcal{R}_{a}+Y^{N+1}+Y^{2}\mathcal{R}_{a}^{2}\mathcal{R}_{b}^{6N-3}+Y^{2}\mathcal{R}_{a}\mathcal{R}_{b}^{6N-2}+\mathcal{R}_{b}^{6N}+\mathcal{R}_{b}^{6N-1})].
\end{array}
$$ The proposition follows readily from this estimate, the easy
inequality
$$
\int^{t}_{0}(t-s)^{-1/2}(T_{0}+s)^{-1/2-\delta}ds\leq
c(T_{0}+t)^{-\delta}$$ valid for any $t\geq 0$ and $1/2>\delta>0,$
and the definition of $\mathcal{R}_{b}$ in Equation
(~\ref{majorant1}).
\end{proof}
\subsection{Proof of Main Theorems ~\ref{maintheorem1} and ~\ref{maintheorem2} for $d=1$}
By Proposition ~\ref{generalestimate} we have that if
$T_{0}^{\frac{2N+1}{2N}}\|\rho_{-2}R(0)\|_{\mathcal{H}^{1}}$ is
sufficiently small and $|Y(0)|$ is bounded, then
$\mathcal{R}_{n}(T),\ n=a,b,c,\ Y(T)\leq c$ for any time $T$. The
rest of the proof of Theorems ~\ref{maintheorem1} and
~\ref{maintheorem2} is just repeat of the proof for the $d\geq 3$
case, given in Subsection ~\ref{subsec:proofsubmaintheorem}.
\begin{flushright}
$\square$
\end{flushright}
\appendix
\section{Proof of Lemma ~\ref{expandk}}\label{AP:Lemma}
Since the proof is long, we begin with Equations
(~\ref{yk})-(~\ref{gammak}) first. Recall that $k>N$ in Lemma
~\ref{expandk}.
\begin{lemma}\label{LM:Sign}
\begin{enumerate}
\item[(1)] The linear functionals $l_{\lambda}^{(k)},\ l_{\gamma}^{(k)}$
and $l_{y}^{(k)}$ satisfy the estimates (~\ref{eq:linEstK}).
\item[(2)]
If the functions $R_{m_{1},n_{1}}(\lambda)$ are admissible for all
pairs $(m_{1},n_{1})<(m,n)$ with $m,n\leq N$ and $m+n\leq k$, then
$\Lambda_{m,n}(\lambda)$ and $\Theta_{m,n}(\lambda)$ in Equations
(~\ref{yk}) and (~\ref{lambdak}) are purely imaginary, and
$\Gamma_{m,n}(\lambda)$ in Equation (~\ref{gammak}) are real.
\item[(3)] $\Theta_{m,n}(\lambda)=Y_{m,n}(\lambda)$ if $m+n\leq N$, where,
recall the definition of $Y_{m,n}(\lambda)$ and the property that
$Y_{m,n}(\lambda)=0$ if $m+n\leq N$ and $m\not=n+1$ in
(~\ref{eq:tranform}).
\end{enumerate}
\end{lemma}
\begin{proof}
\begin{enumerate}
\item[(1)]The estimates on $l_{\lambda}^{(k)},\
l_{\gamma}^{(k)}$ and $l_{y}^{(k)}$ is easy to get by Equations
(~\ref{eq:lambda}) and (~\ref{eq:z1z2}) and the observations that
the functions $\xi,\eta,\phi^{\lambda},\phi^{\lambda}_{\lambda}$
decay exponentially fast.
\item[(2)]
The proof of the properties of $\Lambda_{m,n}(\lambda)$ and
$\Gamma_{m,n}(\lambda)$ are almost the same to those in the proof of
Proposition ~\ref{firststep}, namely in all the computations only
multiplications are involved, thus all pairs $(m,n)$ depend only on
the pairs $(m', n')<(m,n).$
\end{enumerate}
Now we turn to the proof for $\Theta_{m,n}(\lambda).$ Using Equation
(~\ref{eq:z2}) we obtain
\begin{equation}\label{zk}
\dot{z}=i\epsilon(\lambda)z+\sum_{2\leq m+n\leq
2N+1}Y_{m,n}^{(2)}(\lambda)y^{m}\bar{y}^{n}+l_{y}^{(k)}(R_{k})+Remainder
\end{equation} where $l_{y}^{(k)}$ satisfy the estimate (~\ref{eq:linEstK}), and $Remainder$ satisfies Estimate (~\ref{remainder}). By the same arguments as was used for $\Lambda^{(2)}_{m,n}(\lambda)$ we obtain that
$Y_{m,n}^{(2)}(\lambda),$ with $m,n\leq N$ and $m+n\leq k,$ is
purely imaginary if $R_{m',n'}(\lambda)$ are admissible for all the
pairs $(m',n')<(m,n)$.

We invert the function $y=z+P(z,\bar{z})$ in Proposition
~\ref{PRO:transform} to get
\begin{equation}\label{InvertZY}
z=y+\displaystyle\sum_{2\leq m+n\leq
2N+1}P_{m,n}^{(2)}(\lambda)y^{m}\bar{y}^{n}+Remainder
\end{equation}
with $P_{m,n}^{(2)}(\lambda)$ being real. Plug this expression into
Equation (~\ref{zk}) to obtain
$$
\dot{y}=i\epsilon(\lambda)y+D_{1}+D_{2}+l_{y}^{(k)}(R_{k})+Remainder.
$$ with $$
\begin{array}{lll}
D_{1}&:=&\displaystyle\sum_{2\leq m+n\leq
2N+1}Y_{m,n}^{(2)}(\lambda)y^{m}\bar{y}^{n}-\dot{\lambda}\sum_{2\leq
m+n\leq
2N+1}\partial_{\lambda}P_{m,n}^{(2)}(\lambda)y^{m}\bar{y}^{n}\\
& &-i\epsilon(\lambda)(m-n-1)\displaystyle\sum_{2\leq m+n\leq
2N+1}P_{m,n}^{(2)}(\lambda)y^{m}\bar{y}^{n}
\end{array}
$$ and
$$D_{2}:=-\displaystyle\sum_{2\leq m+n\leq
2N+1}P_{m,n}^{(2)}(\lambda)[\frac{d}{dt}y^{m}\bar{y}^{n}-i\epsilon(\lambda)(m-n)y^{m}\bar{y}^{n}].$$

Using (~\ref{lambdak}), which is proved above, for the time
derivatives in the expression for $D_{1}$ we obtain
$$D_{1}=\sum_{2\leq m+n\leq
2N+1}D_{m,n}y^{m}\bar{y}^{n}+l_{y}^{(k)}(R_{k})+Remainder,$$ where
the functionals $l_{y}^{(k)}$ satisfy the estimate
(~\ref{eq:linEstK}). We have that if the functions
$R_{m_{1},n_{1}}(\lambda)$ are admissible for all pairs
$(m_{1},n_{1})<(m,n)$ with $m,n\leq N$ and $m+n\leq k$, then
$D_{m,n}$ are purely imaginary. Indeed, this follows from the
properties of the expansion for $\dot\lambda$ in (~\ref{lambdak}),
which is proved above, and by the properties of
$P^{(2)}_{m,n}(\lambda)$ and $Y^{(2)}_{m,n}(\lambda)$ which we just
mentioned (we omit the detail here).

Substitute in the expression for $D_{2}$ the equation (~\ref{yk}) to
get
\begin{equation}\label{messyterm}
\begin{array}{lll}
D_{2}&=&\displaystyle\sum_{2\leq m'+n'\leq
2N+1,}\sum_{2\leq l_{1}+l_{2}\leq 2N+1}m'P_{m',n'}^{(2)}(\lambda)\Theta_{l_{1},l_{2}}(\lambda)y^{m'+l_{1}-1}\bar{y}^{n'+l_{2}}\\
& &+\displaystyle\sum_{2\leq m'+n'\leq 2N+1,}\sum_{2\leq
l_{1}+l_{2}\leq
2N+1}n'P_{m',n'}^{(2)}\bar{\Theta}_{l_{1},l_{2}}(\lambda)y^{m'+l_{2}}\bar{y}^{n'+l_{1}-1}\\
& &+Remainder.
\end{array}
\end{equation}
We have that if $m'+l_{1}-1,n'+l_{2}\leq N$ then
$$\text{either}\ m'P_{m',n'}^{(2)}\Theta_{l_{1},l_{2}}(\lambda)=0\ \text{or}\
(m'+l_{1}-1,n'+l_{2})>(l_{1},l_{2}),$$ where, recall that
$P_{m',n'}^{(2)}$ are real in (~\ref{InvertZY}). Thus if
$m'+l_{1}-1,\ n'+l_{2}\leq N$ then
$m'P_{m',n'}^{(2)}(\lambda)\Theta_{l_{1},l_{2}}(\lambda)$ is purely
imaginary if $\Theta_{l_{1},l_{2}}$ is purely imaginary for all
pairs $(l_{1},l_{2})<(m'+l_{1}-1,n'+l_{2})$. The same results hold
also for $n'P_{m',n'}^{(2)}\bar{\Theta}_{l_{1},l_{2}}(\lambda).$
Thus if we expand
$$D_{2}=\sum_{2\leq m+n\leq
2N+1}Y^{(4)}_{m,n}(\lambda)y^{m}\bar{y}^{n}+Remainder,$$ then
$Y^{(4)}_{m,n}(\lambda),$ $m,n\leq N,$ are purely imaginary if
$\Theta_{m',n'}(\lambda)$ are purely imaginary for all pairs
$(m',n')<(m,n)$,.

By the discussion above we see that
$\Theta_{m,n}(\lambda)=D_{m,n}+Y^{(4)}_{m,n},$ $m,n\leq N$, $m+n\leq
k,$ are purely imaginary provided that for all pairs $(m',n')<(m,n)$
the functions $R_{m',n'}(\lambda)$ are admissible and
$\Theta_{m',n'}(\lambda)$ are purely imaginary.

Recall the definition and property of $Y_{m,n}(\lambda)$ in
(~\ref{eq:tranform}). We observe that
$Y_{m,n}(\lambda)=\Theta_{m,n}(\lambda)$ when $m+n\leq N$ by the
fact that the expansion in the $k\geq N+1$ step does not affect the
coefficients of $y^{m}\bar{y}^{n},$ $m+n\leq N.$
\end{proof}
Now we turn to the proof of the rest of Lemma ~\ref{expandk}, i.e.
the claim on the function $f_{m,n}$. We plug the equation
(~\ref{expansionR1}) into Equation (~\ref{eq:R1R22}), and use that
$$P_{c}J\vec{N}(\vec{R},z)=\sum_{2\leq m+n\leq
k}y^{m}\bar{y}^{n}P_{c}N_{mn}(\lambda)+A_{k}(y,\bar{y})R_{k}+N_{N}(R_{N},y,\bar{y})+F_{k}(y,\bar{y}),$$
where the terms $A_{k}(y,\bar{y})$, $N_{N}(R_{N},y,\bar{y})$ and
$F_{k}(y,\bar{y})$ are described in Theorem ~\ref{expansion}, and
$k>N$. The result is
$$
\frac{d}{dt}R_{k}=[L(\lambda)+\dot\gamma P_{c}J+\dot\lambda
P_{c\lambda}+A_{k}(y,\bar{y})]R_{k}+\displaystyle\sum_{n=1}^{5}G_{n}+N_{N}(R_{N},y,\bar{y})+F_{k}(y,\bar{y})
$$ where
$$G_{1}:=P_{c}\displaystyle\sum_{2\leq m+n\leq k}y^{m}\bar{y}^{n}[L(\lambda)-i(m-n)\epsilon(\lambda)]R_{m,n}(\lambda);$$
$$G_{2}:=-P_{c}\displaystyle\sum_{2\leq m+n\leq k}[\frac{d}{dt}y^{m}\bar{y}^{n}-i(m-n)\epsilon(\lambda)y^{m}\bar{y}^{n}]R_{m,n}(\lambda);$$
$$G_{3}:=-P_{c}\dot\lambda\displaystyle\sum_{2\leq m+n\leq k}y^{m}\bar{y}^{n}\partial_{\lambda}R_{m,n}(\lambda);$$
$$
G_{4}:=P_{c}\sum_{2\leq m+n\leq k}y^{m}\bar{y}^{n}(\dot\lambda
P_{c\lambda}+\dot\gamma P_{c} J)R_{m,n}(\lambda)
$$ and $$
G_{5}:=\frac{1}{2}\dot\gamma P_{c}[z\left(
\begin{array}{lll}
-i\eta\\
\xi
\end{array}
\right)+\bar{z}\left(
\begin{array}{lll}
i\eta\\
\xi
\end{array}
\right)]-\frac{1}{2}\dot\lambda P_{c}[z\left(
\begin{array}{lll}
\xi_{\lambda}\\
-i\eta_{\lambda}
\end{array}
\right)+\bar{z}\left(
\begin{array}{lll}
\xi_{\lambda}\\
i\eta_{\lambda}
\end{array}
\right)].
$$

Plug the expansions for $\dot{y},$ $\dot\lambda$ and $\dot\gamma$
given in Equations (~\ref{yk})-(~\ref{gammak}), which are proved in
Lemma ~\ref{LM:Sign}, into $G_{l},$ $l=3,\cdot\cdot\cdot,5,$ to
obtain
$$G_{l}=\sum_{2\leq m+n\leq k}y^{m}\bar{y}^{n}G_{m,n}^{(l)}(\lambda)+F_{k}(y,\bar{y})$$
where for each $m,n\leq N$ the function $iG^{(l)}_{mn}(\lambda)$ is
admissible if $R_{m',n'}(\lambda)$ are admissible for all pairs
$(m',n')<(m,n).$ Moreover, if $R_{m',n'}(\lambda)$ are of the forms
$$\prod_{k}(L(\lambda)-ik\epsilon(\lambda)+0)^{-n_{k}}P_{c}\phi_{m',n'}(\lambda)$$ for
all the pairs $(m',n')<(m,n)$ then, using the observation that
$P_{c}\partial_{\lambda}\prod_{k}(L(\lambda)-ik\epsilon(\lambda)+0)^{-n_{k}}P_{c}\phi_{m',n'}(\lambda)$
and
$P_{c}J\prod_{k}(L(\lambda)-ik\epsilon(\lambda)+0)^{-n_{k}}P_{c}\phi_{m',n'}(\lambda)$
are of the form
$\prod_{k}(L(\lambda)-ik\epsilon(\lambda)+0)^{-n_{k}}P_{c}\phi^{(2)}_{m',n'}(\lambda),$
we can show that the functions $G^{(l)}_{m,n}(\lambda),$
$\max\{m,n\}>N,$ are of a similar form also.

For $G_{2}$, using the equation for $y$ in (~\ref{yk}) we have $$
\begin{array}{lll}
G_{2}&=&\displaystyle\sum_{2\leq m'+n'\leq k}\sum_{2\leq m+n\leq k}mR_{mn}(\lambda)\Theta_{m',n'}(\lambda)y^{m-1+m'}\bar{y}^{n+n'}\\
& &+\displaystyle\sum_{2\leq m'+n'\leq k}\sum_{2\leq m+n\leq
k}nR_{m,n}(\lambda)\bar{\Theta}_{m',n'}(\lambda)y^{m+n'}\bar{y}^{n-1+m'}+F_{k}(y,\bar{y}).
\end{array}
$$ Recall the definition
and property of $\Theta_{m,n}(\lambda)$ in (~\ref{yk}), we have that
if $m-1+m',n+n'\leq N,$ then
$$\text{either}\ mR_{m,n}(\lambda)\Theta_{m',n'}(\lambda)=0\ \text{or}\
(m,n)<(m-1+m',n+n')$$ (this is the point where we use the property
that $\Theta_{m,n}(\lambda)=0$ for $m,n\leq N$ and $m\not=n+1$,
where, recall the fact that $\Theta_{m,n}(\lambda)=Y_{m,n}(\lambda)$
if $m+n\leq N$ proved in Lemma ~\ref{LM:Sign}). Thus
$$G_{2}=\sum_{2\leq m+n\leq k}G^{(2)}_{m,n}(\lambda)y^{m}\bar{y}^{n}+F_{k}(y,\bar{y}),$$
where if $m,n\leq N$ and $R_{m',n'}(\lambda)$ are admissible for all
the pairs $(m',n')<(m,n)$ then $iG_{m,n}^{(2)}(\lambda)$ is
admissible.

This completes the proof of Lemma ~\ref{expandk}.
\begin{flushright}
$\square$
\end{flushright}
\section{Transformation of $y$}\label{AP:Transform}
In this appendix we prove a result used in the proof of Lemma
~\ref{transformz}.
\begin{proposition}\label{Co:invariant}
Let complex and real functions $y(t)$ and $\lambda(t)$ satisfy
Equations (~\ref{expanz2})-(~\ref{remainder}) and
(~\ref{expanlambda}), and let $P(y,\bar{y})$ be a polynomial of the
form
$$P(y,\bar{y})=\displaystyle\sum_{N+1\leq m+n\leq
2N+1}p_{m,n}(\lambda)y^{m}\bar{y}^{n}$$ with the coefficients
$p_{m,n}(\lambda)$ real for $m,n\leq N$. Define
$\beta:=y+P(y,\bar{y}).$ Then we have
\begin{equation}\label{eq:invarlambda}
\dot{\lambda} =\sum_{2\leq m+n\leq
2N+1}\Lambda_{m,n}^{(1)}(\lambda)\beta^{m}\bar{\beta}^{n}+Remainder
\end{equation}
and
\begin{equation}\label{eq:invarbeta}
\dot{\beta}=i\epsilon(\lambda)\beta+\sum_{3\leq m+n\leq 2N+1
}\Theta^{(1)}_{m,n}(\lambda)\beta^{m}\bar{\beta}^{n}+Remainder
\end{equation}
where the functions $\Lambda^{(1)}_{m,n}(\lambda)$ and
$\Theta^{(1)}_{m,n}(\lambda)$ are purely imaginary for $m,n\leq N;$
$\Theta^{(1)}_{m,n}=0$ if $m+n\leq N$ and $m\not= n+1;$ and the term
$Remainder$ stands for a function satisfying (~\ref{remainder}).
\end{proposition}
\begin{proof}
We invert the relation $\beta:=y+P(y,\bar{y})$ to get the expression
\begin{equation}\label{yinbeta}
y=\beta+\displaystyle\sum_{N+1\leq m+n\leq
2N+1}P_{m,n}^{(2)}(\lambda)\beta^{m}\bar{\beta}^{n}+O(|\beta|^{2N+2}).
\end{equation}
Since $p_{m,n}(\lambda)$ are real for $m,n\leq N$, the coefficients
$P_{m,n}^{(2)}(\lambda)$ are real for $m,n\leq N$.

Plug Equation (~\ref{yinbeta}) into Equation (~\ref{expanlambda}) to
obtain
$$
\dot{\lambda} =\displaystyle\sum_{2\leq m+n\leq
2N+1}\Lambda^{(1)}_{m,n}(\lambda)\beta^{m}\bar{\beta}^{n}+Remainder.
$$ We claim that $\Lambda^{(1)}_{m,n}(\lambda)$ are purely
imaginary for $m,n\leq N$. Indeed, we observe that
$$\Lambda^{(1)}_{m,n}(\lambda)=\Lambda_{m,n}(\lambda)+\sum_{
\begin{subarray}{lll}
m'+l_{1}=m+1\\
n'+l_{2}=n
\end{subarray}
}m'\Lambda_{m',n'}P^{(2)}_{l_{1},l_{2}}(\lambda)+\sum_{
\begin{subarray}{lll}
m'+l_{2}=m\\
n'+l_{1}=n+1
\end{subarray}
}n'\Lambda_{m',n'}\bar{P}^{(2)}_{l_{1},l_{2}}(\lambda),$$ where,
recall, $\Lambda_{m,n}(\lambda)$ are purely imaginary for $m,n\leq
N.$ Since $l_{1}+l_{2}\geq N+1$, we have that if $m,n\leq N,$
$m'\not=0,$ $m'+l_{1}=m+1$ and $n'+l_{2}=n$ then
$m',n',l_{1},l_{2}\leq N.$ This implies that
$\Lambda_{m',n'}(\lambda)$ are purely imaginary and
$P^{(2)}_{l_{1},l_{2}}(\lambda)$ are real. Hence
$m'\Lambda_{m',n'}P^{(2)}_{l_{1},l_{2}}(\lambda)$ are purely
imaginary if $m'\not=0$ (If $m'=0$ then
$m'\Lambda_{m',n'}P^{(2)}_{l_{1},l_{2}}(\lambda)=0$). By the same
reasoning we prove that
$n'\Lambda_{m',n'}\bar{P}^{(2)}_{l_{1},l_{2}}(\lambda)$ is purely
imaginary for $m,n\leq N$. These two facts together with
$\Lambda_{m,n}(\lambda)$ being purely imaginary for $m,n\leq N$
imply that $\Lambda^{(1)}_{m,n}(\lambda)$ are purely imaginary for
$m,n\leq N.$ This completes the proof of Equation
(~\ref{eq:invarlambda}) and its properties.

Now we turn to Equation (~\ref{eq:invarbeta}). By Equation
(~\ref{expanz2}) we obtain
\begin{equation}\label{betatransform}
\begin{array}{lll}
\dot{\beta}&=&i\epsilon(\lambda)\beta+\displaystyle\sum_{2\leq
m+n\leq 2N+1}\Theta_{mn}(\lambda)y^{m}\bar{y}^{n}+K +Remainder,
\end{array}
\end{equation} where the term $K$ is defined as $$K:=\frac{d}{dt}\displaystyle\sum_{N+1\leq m+n\leq
2N+1}p_{m,n}(\lambda)y^{m}\bar{y}^{n}-i\epsilon(\lambda)\displaystyle\sum_{N+1\leq
m+n\leq 2N+1}p_{m,n}(\lambda)y^{m}\bar{y}^{n},$$ and recall, the
coefficients $\Theta_{m,n}$ are defined in  (~\ref{expanz2}).

Using Equation (~\ref{expanz2}) we obtain
\begin{equation}\label{expandy}
K=\displaystyle\sum_{N+1\leq m+n\leq
2N+1}P_{m,n}(\lambda)y^{m}\bar{y}^{n}+Remainder.
\end{equation} We show below that $P_{m,n}(\lambda)$ are purely imaginary for
$m,n\leq N;$ and $P_{m,n}(\lambda)=0$ for $m+n\leq N.$ This fact
implies that Equation (~\ref{betatransform}) is of the form
\begin{equation}\label{step1}
\dot{\beta}=i\epsilon(\lambda)\beta+\displaystyle\sum_{2\leq m+n\leq
2N+1}\Theta^{(2)}_{mn}(\lambda)y^{m}\bar{y}^{n}+Remainder,
\end{equation} where $\Theta^{(2)}_{m,n}(\lambda)$ are purely imaginary
for $m,n\leq N,$ and $\Theta^{(2)}_{m,n}=0$ if $m+n\leq N$ and
$m\not= n+1.$ Substitute into the right hand side of the expansion
for $y$ given by Equation (~\ref{yinbeta}) to obtain a new equation
for $\beta$
\begin{equation}\label{step2}
\begin{array}{lll}
\dot{\beta}&=&i\epsilon(\lambda)\beta+\displaystyle\sum_{2\leq
m+n\leq
2N+1}\Theta^{(2)}_{m,n}(\lambda)\beta^{m}\bar{\beta}^{n}\\
& &+\displaystyle\sum_{ N+1\leq m'+n'}\sum_{2\leq m+n\leq
2N+1}m\Theta^{(2)}_{m,n}(\lambda)P_{m',n'}^{(2)}\lambda\beta^{m+m'-1}\bar{\beta}^{n+n'}\\
& &+\displaystyle\sum_{ N+1\leq m'+n'}\sum_{2\leq m+n\leq
2N+1}n\Theta^{(2)}_{m,n}(\lambda)\bar{P}_{m',n'}^{(2)}(\lambda)\beta^{m+n'}\bar{\beta}^{n+m'-1}+Remainder\\
&=&i\epsilon(\lambda)\beta+\displaystyle\sum_{2\leq m+n\leq
2N+1}\Theta^{(3)}_{m,n}(\lambda)\beta^{m}\bar{\beta}^{n}.
\end{array}
\end{equation}
By the properties of $\Theta^{(2)}_{m,n}$ above and the facts that
$P_{m,n}^{(2)}(\lambda)$ are real if $m,n\leq N$ and
$P_{m,n}^{(2)}(\lambda)=0$ if $m+n\leq N$, we have that
$\Theta^{(3)}_{m,n}(\lambda)=0$ if $m\not=n+1$ and $m+n\leq N;$
$\Theta^{(3)}_{m,n}(\lambda)$ is purely imaginary for $m,n\leq N.$

To complete the proof of Proposition ~\ref{Co:invariant} it remains
to prove the claim above that $P_{m,n}(\lambda)$ are purely
imaginary for $m,n\leq N$ and $P_{m,n}(\lambda)=0$ for $m+n\leq N$.
Compute
\begin{equation}\label{fourterms}
\begin{array}{lll}
K &=&\dot\lambda\displaystyle\sum_{N+1\leq m+n\leq
2N+1}\partial_{\lambda}p_{m,n}(\lambda)y^{m}\bar{y}^{n}\\
& &+\displaystyle\sum_{N+1\leq m+n\leq 2N+1}
p_{m,n}(\lambda)[\frac{d}{dt}y^{m}\bar{y}^{n}-i\epsilon(\lambda)y^{m}\bar{y^{n}}]\\
&=&\displaystyle\sum_{2\leq m'+n'\leq 2N+1,}\sum_{N+1\leq m+n\leq 2N+1}\partial_{\lambda}p_{m,n}(\lambda)\Lambda_{m',n'}y^{m+m'}\bar{y}^{n+n'}\\
& &+\displaystyle\sum_{N+1\leq m+n\leq 2N+1,}\sum_{2\leq m'+n'\leq
2N+1}mp_{m,n}(\lambda)\Theta_{m',n'}y^{m-1+m'}\bar{y}^{n+n'}\\
& &+\displaystyle\sum_{N+1\leq m+n\leq 2N+1,}\sum_{2\leq m'+n'\leq
2N+1}np_{m,n}\bar{\Theta}_{m',n'}(\lambda)y^{m+n'}\bar{y}^{n-1+m'}\\
& &+i\epsilon(\lambda)\displaystyle\sum_{N+1\leq m+n\leq
2N+1}(m-n-1)p_{m,n}(\lambda)y^{m}\bar{y}^{n}+Remainder.
\end{array}
\end{equation}
We have that $P_{m,n}(\lambda)=0$ for $m+n\leq N$ since all the
expressions above are of order $o(|y|^{N}).$ Next we show that
$P_{m,n}(\lambda)$ are purely imaginary for $m,n\leq N.$ We have the
following observations for the four terms on the right hand side of
(~\ref{fourterms})
\begin{enumerate}
\item[(A)] if $m+m', n+n'\leq N,$
then we have $m,n,m',n'\leq N$ which implies that
$\Theta_{m',n'}(\lambda)$ is purely imaginary and $p_{m,n}(\lambda)$
is real. Thus $\partial_{\lambda}p_{m,n}\Theta_{m',n'}(\lambda)$ is
purely imaginary;
\item[(B)] if $m-1+m', n+n'\leq N,$ then either
$mp_{m,n}\Theta_{m',n'}(\lambda)$ is zero or $m,n,m',n'\leq N$ by
the properties of $p_{m,n}(\lambda)$ and $\Theta_{m',n'}(\lambda)$.
Thus $mp_{m,n}\Theta_{m',n'}(\lambda)$ is purely imaginary;
\item[(C)] if $m+n', n-1+m'\leq N,$ then
$np_{m,n}\bar{\Theta}_{m',n'}(\lambda)$ is purely imaginary by the
same reasoning as in (B) above;
\item[(D)]$i\epsilon(\lambda)p_{m,n}(\lambda)$ is purely imaginary for $m,n\leq N$
since the coefficients $p_{m,n}(\lambda)$ are real.
\end{enumerate}

Collecting the results above we conclude that $P_{m,n}(\lambda)$ are
purely imaginary for $m,n\leq N$. This completes the proof of the
claim made after Equation (~\ref{expandy}) and, with it, the proof
of Proposition ~\ref{Co:invariant}.
\end{proof}

\end{document}